\documentclass[reprint, 
nofootinbib,
 amsmath,amssymb,
 aps, physrev,
floatfix,
]{revtex4-2}

\usepackage[caption=false]{subfig}
\usepackage{graphicx}
\usepackage{dcolumn}
\usepackage{subfig}
\usepackage{bm}
\usepackage{enumitem}
\usepackage{hyperref}

\usepackage[english]{babel}

\usepackage{array}
\usepackage{booktabs} 

\usepackage{soul, xcolor}
\sethlcolor{yellow} 

\newcolumntype{L}[1]{>{\raggedright\arraybackslash}p{#1}}

\begin{document}

\preprint{APS/123-QED}

\title{\textbf{Dynamical Lie Algebras Cannot Describe Shallow QAOA: Cragged Terrains, Barren Plateaus, and Empirical Hardness Models} 
}

\author{Harrison Copp}
\affiliation{Yale University}
\altaffiliation{Author order is alphabetical.}

\author{Charlton Li}
\affiliation{The Ohio State University}

\author{An\v{z}ej Margeta-Cacace}
\email{anzejmc@gmail.com}
\affiliation{Texas Tech University}

\author{Amy Qiao}
\affiliation{Brown University}
\date{\today}

\begin{abstract}
The dynamical Lie algebraic (DLA) theory of variational quantum algorithms (VQAs) predicts commonplace exponentially vanishing loss and gradient variances for sufficiently deep parametrized circuits. In this work, we show that these predictions fail dramatically in the shallow-circuit (and particularly constant-depth) regime for the Quantum Approximate Optimization Algorithm (QAOA) applied to the maximum independent set (MIS) problem. In a large-scale numerical study across $\sim$23,000 problem instances, we find that barren plateaus are rare, while landscapes whose variances polynomially \emph{increase} with system size---which we term “cragged terrains”---are common across graph families. This aggregate polynomial growth persists both for generic, low-symmetry random graphs and for highly symmetric vertex-transitive graphs, indicating that DLA-based variance predictions do not describe landscape scaling in this regime. As a stopgap alternative to the theory, we train empirical hardness models to predict instance-wise hardness metrics for QAOA-MIS. While these models generalize poorly, they nonetheless recover the correct landscape scaling class (barren plateau vs. cragged terrain) with high fidelity. Taken together, our results identify shallow QAOA for MIS as a prototypical setting in which asymptotic, unitary-design-centric predictions may be fundamentally insufficient to describe shallow variational quantum algorithms more broadly, emphasizing the need for more empirically-informed models of VQA loss landscapes. 
\end{abstract}

\maketitle


\section{Introduction}

Variational quantum algorithms (VQAs) leverage classical optimization techniques to optimize a parametrized quantum circuit. As this optimization process keeps circuit depths low, VQAs have become a promising candidate to operate on near-term quantum hardware \cite{cerezo_variational_2021, larocca_barren_2025}. A significant bottleneck to implementing VQAs, however, is the so-called ``barren plateau'' phenomenon, where the loss landscape becomes exponentially flat as the number of qubits increase \cite{mcclean_barren_2018, larocca_barren_2025, cerezo_variational_2021}. As VQAs rely on a finite number of measurements to estimate the loss, exponentially concentrated loss values in general require an exponential number of measurements to successfully navigate the loss landscape \cite{mcclean_barren_2018, larocca_barren_2025}. Barren plateaus therefore have the potential to nullify any potential quantum advantage. 

Recently, a robust theoretical framework for predicting the presence of barren plateaus has emerged \cite{mcclean_barren_2018, cerezo_cost_2021, ragone_lie_2024, fontana_adjoint_2024, diaz_showcasing_2023, anschuetz_unified_2025, larocca_diagnosing_2022}. By assuming that the quantum circuit forms an approximate unitary 2-design over an ambient Lie group, recent work has derived closed-form expressions for the variance of the loss landscape and loss landscape gradients \cite{ragone_lie_2024, fontana_adjoint_2024, diaz_showcasing_2023, anschuetz_unified_2025}. These expressions verify commonly held heuristic knowledge about barren plateau behavior, such as the fact that barren plateaus appear more often with highly expressive parametrized quantum circuits or highly entangled initial states \cite{larocca_barren_2025, ragone_lie_2024, fontana_adjoint_2024}. They also allow for robust theoretical guarantees about when certain VQA instances will or will not encounter barren plateaus \cite{ragone_lie_2024}. As the backbone of the theory was developed using the dynamical Lie algebra (DLA) \cite{ragone_lie_2024, fontana_adjoint_2024}, we refer to this theory collectively as DLA theory. Although more recent results have moved beyond the DLA \cite{anschuetz_quantum_2022, diaz_showcasing_2023}, they also make use of asymptotic algebraic objects to model circuit behavior (even potentially shallow circuit behavior \cite{anschuetz_unified_2025, mao2025qaoamaxcutbarrenplateausgraphs}), rely crucially on design-centric assumptions, and usually reference DLA results as a basis for more generalized formulas.

These results are a promising step forward in understanding barren plateau behavior. However, as the key 2-design assumption is only guaranteed to hold in deeper circuit regimes, it is unclear---and, as we will argue, unlikely---that these results apply to the shallower circuit depths of near-term   
interest \cite{ragone_lie_2024, fontana_adjoint_2024, larocca_barren_2025}. 

These limitations in existing barren plateau theory motivate our empirical approach to understanding barren plateaus in the shallow circuit regime. In particular, we employ \textit{empirical hardness models} (EHMs), which supplement traditional worst-case analysis by training a machine learning model to predict an indicator of problem hardness (e.g., algorithmic runtime) from datasets of typical problem instances \cite{leyton-brown_empirical_2009, leyton-brown_understanding_2014}. In our case, our models learn a relationship between VQA problem instances and loss landscape variance. The promise of EHMs for our purpose is that they may be able to train on inexpensive problem instances (low qubit count, shallow circuits) and extrapolate upwards.

We conduct our empirical analysis using the Quantum Approximate Optimization Algorithm (QAOA) \cite{farhi_quantum_2014}, a standard VQA designed for combinatorial optimization problems, applied to maximum independent set (MIS), the problem of finding the largest set of vertices in a graph such that no two vertices are connected by an edge. We choose MIS because it is a benchmark NP-hard problem that also has salient real world applications, including satellite scheduling \cite{brady_iterative_2023, eddy_maximum_2020}.

In the course of our analysis, we show that EHMs are effective predictors of loss landscape variance, provided that tested problem instances are structurally similar to the training data. Furthermore, the models were capable of capturing the scaling relationship between the number of qubits and the loss landscape variance to a high degree of accuracy, indicating that our models could be deployed to predict barren plateaus. Nonetheless, our models did not generalize well to higher qubit counts or deeper circuits, providing a preliminary negative answer to the promise stated earlier. It is unclear whether the inability to generalize was a problem with our model choice and architecture, or reflective of a quality of the graph distributions on which we trained, where larger graphs (especially for deeper circuits) may exhibit fundamentally different properties to smaller ones.

In the process of collecting training data for our EHMs, we were also able to collect extensive statistics about the quantity and characteristics of barren plateaus in QAOA applied to MIS. In total, we studied all graphs up to 7 vertices, $\sim$3,600 Erdős–Rényi, Watts–Strogatz, and Barabási–Albert random graphs up to 20 vertices, and the complete set of vertex-transitive graphs up to order 20, ranging across circuit depths between 1 and 10, amounting to $\sim$23,000 unique problem instances. We found that barren plateaus were rare, and, moreover, the opposite behavior---a polynomial \textit{increase} in the loss landscape variance---which we term ``cragged terrains’’, were common. Furthermore, in compiling all problem instances, we observed an aggregate polynomial increase, a departure from previous studies \cite{mao2025qaoamaxcutbarrenplateausgraphs, larocca_diagnosing_2022, yao2026gradientanalysisbarrenplateau, Kashif_2024}.

Despite the insinuations of design-centric theories \cite{mao2025qaoamaxcutbarrenplateausgraphs, anschuetz_quantum_2022}, barren plateau behavior is the exception rather than the rule in the shallow-circuit regime: cragged terrains dominate, and the aggregate variance grows with system size across graph families as structurally diverse as generic random graphs and highly symmetric vertex-transitive graphs alike. This suggests that design-centric theories may be fundamentally ill-suited to describe shallow VQA landscapes.

We structure this paper as follows. In Section \ref{sec:background}, we discuss  prerequisite information about variational quantum algorithms, barren plateaus, and the maximum independent set problem. In Subsections \ref{sec:unitary_designs} and \ref{sec:bp_theory}, we conduct a review of current theoretical barren plateau analysis and establish how the widely-assumed approximate 2-design conditions limits the applicability of this theory to shallow circuits. In Section \ref{sec:methods}, we discuss our how we built our graph dataset, how we identified barren plateaus, and the architectures of our EHMs. In Section \ref{sec:results}, we present our EHM model performance and the results of our numerical analysis, noting the rarity of barren plateaus and preponderance of cragged terrains for shallow QAOA MIS, as well as EHMs' superb effectiveness at classifying asymptotic landscape scaling behavior. Finally, in Section \ref{sec:discussion}, we discuss the theoretical implications of our numerical results and propose future directions for research.

\section{Background}
\label{sec:background}

\subsection{Variational Quantum Algorithms}

Suppose we are given a problem instance with an associated cost function $C(\boldsymbol{\theta})$, such that the solution is encoded by $\boldsymbol{\theta}^* = \arg\min C(\boldsymbol{\theta})$. This task can be approached using any host of classical optimization techniques. However, in cases where evaluating $C(\boldsymbol{\theta})$ is costly, variational quantum algorithms offer a potential speedup by using quantum computing to efficiently evaluate $C(\boldsymbol{\theta})$ while using classical optimization to update the parameters $\boldsymbol{\theta}$ \cite{cerezo_variational_2021}.

More precisely, variational quantum algorithms consist of a parametrized quantum circuit (PQC) ansatz, which we write as $U(\boldsymbol{\theta})$. We take our parameters to be a vector of $L$ real numbers, $\boldsymbol{\theta} = (\theta_1,\cdots,\theta_L)\in \mathbb{R}^L$, so we can view the PQC as a map $U(\boldsymbol{\theta}) : \mathbb{R}^L\to \mathcal{U}(2^N)$, where $\mathcal{U}(2^N)$ is the $2^N\times2^N$ complex unitary matrices operating over an $N$ qubit Hilbert space $\mathcal{H}=(\mathbb{C}^2)^{\otimes N}\cong \mathbb{C}^{2^N}$. The form of the ansatz can be problem-specific, though for purposes of this paper, we consider PQCs of the general form:
\begin{equation}\label{eq:general_pqc}
    U(\boldsymbol{\theta}) = e^{-i\theta_LH_L}e^{-i\theta_{L-1}H_{L-1}}\cdots e^{-i\theta_1H_1}
\end{equation}
where $H_i\in i\mathfrak{u}(2^N)$ are Hermitian operators. We refer to the integer $L$ as the circuit depth.

Now, given a starting state $\rho$ in the Hilbert space $\mathcal{H}=(\mathbb{C}^2)^{\otimes N}$ we may try to express the cost function as:
\begin{equation}
\label{eq:cost-function-general}
    C(\boldsymbol{\theta}) = \mathrm{Tr}[U(\boldsymbol{\theta})\rho U^\dagger(\boldsymbol{\theta})O]
\end{equation}
for some Hermitian observable $O\in i\mathfrak{u}(2^N)$. If such an observable exists, then evaluating the cost function $C(\boldsymbol{\theta})$ is as simple as repeatedly measuring the observable $O$. Once the cost function is evaluated, classical optimization techniques (e.g. gradient descent) can be used to update the parameters $\boldsymbol{\theta}$.

\subsection{The Quantum Approximate Optimization Algorithm}

The Quantum Approximate Optimization Algorithm (QAOA) \cite{farhi_quantum_2014} is a particular type of VQA often used to find approximate solutions to combinatorial optimization problems. Problems like Max Independent Set (MIS), which can be formulated as quadratic unconstrained binary optimization (QUBO) problems \cite{glover2019tutorialformulatingusingqubo} (cf. Section \ref{sec:mis}), are well-suited for QAOA as usually only quantum gates on one or two qubits at a time need to be implemented on hardware.

The QAOA ansatz is given by the following form:
\begin{equation}\label{eq:qaoa_ansatz}
    U(\boldsymbol{\gamma}, \boldsymbol{\beta}) = e^{-i\beta_p\hat{B}}e^{-i\gamma_p\hat{C}}\cdots e^{-i\beta_1\hat{B}}e^{-i\gamma_1\hat{C}}
\end{equation}
where $\boldsymbol{\gamma}$ and $\boldsymbol{\beta} \in [0, 2\pi)^p$ are the parameters. Note the QAOA ansatz follows the general form given by Equation \ref{eq:general_pqc}, with $\boldsymbol{\theta} = \boldsymbol{\gamma},\boldsymbol{\beta}$ and $L=2p$.

As for the Hermitian operators $\hat{B}$, the mixer Hamiltonian, we take:
\begin{equation}
    \hat{B}=-\sum_{i=1}^N\sigma_i^{(x)}
\end{equation}
and $\hat{C}$ depends on the optimization problem (see Section \ref{sec:mis}) and encodes the desired solution states as the set of eigenvectors with the lowest eigenvalue. We initialize the circuit in the state of uniform superposition:
\begin{equation}
    |s\rangle=\frac{1}{\sqrt{2^N}}\sum_{\boldsymbol{x}\in\{0,1\}^N}|\boldsymbol{x}\rangle,
\end{equation}
i.e., the ground state of $\hat{B}$. We write the output of the circuit as $|{\boldsymbol{\gamma}, \boldsymbol{\beta}}\rangle$, where:
\begin{equation}
    |\boldsymbol{\gamma},\boldsymbol{\beta}\rangle= U(\boldsymbol{\gamma}, \boldsymbol{\beta})|s\rangle
\end{equation}
Finally, the cost function retrieving the general form of Equation \ref{eq:cost-function-general} is given by:
\begin{equation}
C(\boldsymbol{\gamma},\boldsymbol{\beta})=\langle\boldsymbol{\gamma},\boldsymbol{\beta}|\,\hat{C}\,|\boldsymbol{\gamma},\boldsymbol{\beta}\rangle.
\end{equation}

\subsection{Barren Plateaus}

In essence, a barren plateau is a feature of how the loss landscape of a VQA scales with respect to the system size (number of qubits). A given VQA is said to exhibit a barren plateau if the loss landscape becomes exponentially flat as the system size increases.

More formally, we say a cost function $C(\boldsymbol{\theta})$ has a barren plateau if the variance of its loss landscape gradients decays exponentially with respect to the number of qubits, that is:
\begin{equation}\label{eq:cost function grad var}
    \mathrm{Var}_{\boldsymbol{\theta}}[\partial_\mu C(\boldsymbol{\theta})] \in \mathcal{O}\Big(\frac{1}{b^N}\Big)
\end{equation}
for all gradient directions $\theta_\mu\in\boldsymbol{\theta}$, where $N$ is the number of qubits and $b>1$ is a real number \cite{mcclean_barren_2018, larocca_barren_2025}.

In general, a loss landscape can show barren plateau behavior along only some, but not all, gradient directions \cite{larocca_barren_2025}. We focus on the case where a barren plateau appears in all directions, as this is what is most often discussed in the literature \cite{cerezo_variational_2021, larocca_barren_2025, mcclean_barren_2018, ragone_lie_2024, fontana_adjoint_2024}. In this case, we can equivalently identify barren plateaus by looking directly at the variance of the loss landscape \cite{larocca_barren_2025, Arrasmith_2022}:
\begin{equation}\label{eq:cost function var}
    \mathrm{Var}_{\boldsymbol{\theta}}[ C(\boldsymbol{\theta})] \in \mathcal{O}\Big(\frac{1}{b^N}\Big)
\end{equation}

This correspondence between the variance of loss function gradients and variance of the loss function in Eq. (\ref{eq:cost function var}) informs us as to why we can speak of barren plateaus as exponentially flat loss landscapes. 

Barren plateaus pose significant problems for VQAs because of their sampling-based approach to computing variance. Successfully navigating the exponentially flat landscapes of a barren plateau requires exponential precision in cost function calculations \cite{larocca_barren_2025, mcclean_barren_2018}. But as the cost function is approximated via a finite number of measurements, this means that barren plateaus demand a number of measurements exponential in $N$ \cite{mcclean_barren_2018, larocca_barren_2025, cerezo_variational_2021}. This could preclude all of the desired quantum advantage of VQAs. Since a barren plateau implicates the accuracy of cost function approximation, it is problematic for gradient-based and gradient-free optimization methods alike \cite{cerezo_variational_2021, larocca_barren_2025, mcclean_barren_2018}.

\subsection{Maximum Independent Set}
\label{sec:mis}

We now define the specific combinatorial optimization problem that we study. Consider a graph $G=(V,E)$ with vertex set $V=\{1,2,\dots,N\}$ and edge set $E$. An \emph{independent set} of $G$ is a subset $S\subseteq V$ such that no two vertices in $S$ share an edge. As the name suggests, the maximum independent set (MIS) problem is to find an independent set of $G$ of largest size.

MIS can be formulated as a quadratic unconstrained binary optimization (QUBO) problem, a class of problems amenable to QAOA \cite{brady_iterative_2023}. Generally, a QUBO problem is one that consists of finding a bitstring $\boldsymbol{x}\in\{0,1\}^N$ that minimizes the cost function
\begin{equation}
    c(\boldsymbol{x})=\boldsymbol{x}^T Q\boldsymbol{x}
\end{equation}
for a given symmetric matrix $Q\in\mathbb{R}^{N\times N}$. In the case of MIS, a bitstring $\boldsymbol{x}=x_1\dots x_N$ corresponds to the vertex subset $\{i\;:\;x_i=1\}\subseteq V$. By relaxing the hard constraint of independence, a cost function for MIS can be written with an edge penalty term:
\begin{align}
    c(\boldsymbol{x})&=-2\sum_{i\in V}x_i+4\lambda\sum_{\{i,j\}\in E}x_i x_j \label{eq:cost_function} \\
    &=\boldsymbol{x}^T (-2I+2\lambda A)\boldsymbol{x},
\end{align}
where $\lambda>0$ is the Lagrange multiplier, and $A$ is the adjacency matrix of $G$. The cost Hamiltonian $\hat{C}$ used by QAOA for MIS is obtained by making the substitution $x_i\to\frac{1}{2}(I-\sigma_i^{(z)})$ to the classical cost function $c(\boldsymbol{x})$ from Eq. \ref{eq:cost_function}, giving
\begin{equation}\label{eq:mis-hamiltonian}
    \hat{C}=-\sum_{i\in V} (I-\sigma_i^{(z)})+\lambda\sum_{\{i,j\}\in E}(I-\sigma_i^{(z)})(I-\sigma_j^{(z)}).
\end{equation}
This substitution ensures that the eigenvalues of $\hat{C}$ correspond to the values of $c$, that $\hat{C}$ is diagonal in the computational basis, and that the eigenvalue of $\hat{C}$ on the quantum state corresponding to a bitstring $\boldsymbol{x}\in\{0,1\}^N$ coincides with the classical cost value $c(\boldsymbol{x})$.

\subsection{Unitary Designs}
\label{sec:unitary_designs}

Since the first paper on barren plateaus \cite{mcclean_barren_2018}, a powerful suite of tools to predict barren plateaus has emerged. However, the theory almost universally assumes asymptotically deep ansätze; more precisely, it requires that the ensemble of unitaries generated by the parametrized quantum circuit forms a unitary 2-design (or an $\varepsilon$-approximate 2-design) \cite{mcclean_barren_2018, larocca_barren_2025, cerezo_cost_2021, ragone_lie_2024, fontana_adjoint_2024, diaz_showcasing_2023, anschuetz_unified_2025}. Unfortunately, the 2-design condition has thus far only been shown to hold for sufficiently deep quantum circuits or specialized circuit architectures \cite{cleve2016nearlinearconstructionsexactunitary}; it is unclear whether shallow quantum circuits form 2-designs \cite{larocca_diagnosing_2022, ragone_lie_2024, fontana_adjoint_2024}. Furthermore, it is expensive to verify whether a given quantum circuit forms a 2-design, precluding any chance of a verify-as-you-go analysis. In this section, we introduce the 2-design condition, analyze its role in barren plateau literature, and finally discuss the practical limitations it imposes on existing barren plateau theory.

We first provide some definitions. An ensemble of unitaries $(\mathcal{E}, \nu)$ is a subset $\mathcal{E}\subseteq\mathcal{U}(d)$ of $d\times d$ unitary matrices along with a measure $\nu$ on $\mathcal{U}(d)$. In particular, we consider the ensemble of unitaries associated with a parametrized quantum circuit $U(\boldsymbol{\theta})$, which is defined as:
\begin{equation}
     \mathcal{E}_L := \left\{U(\boldsymbol{\theta})\in \mathcal{U}(2^N) \mid \boldsymbol{\theta}\in \Theta \right\}
\end{equation}
where $N$ is the number of qubits, $L$ is the circuit depth, and $\Theta$ is the set of all possible parameters quotiented by the appropriate periodic structure corresponding to that of $U(\boldsymbol{\theta})$. Thus the ensemble of unitaries $(\mathcal{E}_L, \nu)$ associated with $U(\boldsymbol{\theta})$ is simply the set of all unitary matrices that can be obtained by varying the parameters $\boldsymbol{\theta}$, coupled with a distribution over those unitaries. 

Now, given an ensemble of unitaries $(\mathcal{E}, \nu)$ and a compact subgroup $G\leq\mathcal{U}(d)$, we say $(\mathcal{E}, \nu)$ forms a 2-design over $G$ if for all degree $2$ balanced monomials $M$ on $\mathcal{U}(d)$, the following equality holds:
\begin{equation}\label{eq:2-design}
    \mathbb{E}_{\nu}[M(U)] = \mathbb{E}_{\mu}[M(U)]
\end{equation}
where $\mu$ is the unique bi-invariant Haar measure on the compact group $G$. A degree 2 balanced monomial $M$ is a monomial that is degree 2 in the matrix elements of a unitary matrix $U$ and degree 2 in the matrix elements of $U^\dagger$. We will also encounter the looser notion of an $\varepsilon$-approximate 2-design, where the equality in Equation \ref{eq:2-design} is true up to $\varepsilon$ with respect to some choice of finite-dimensional matrix norm. Intuitively, if an ensemble $(\mathcal{E}, \nu)$ satisfies the 2-design condition (or more generally, a $t$-design condition), then averaging sufficiently low degree expressions over the ensemble is the same as integrating over $G$ using the Haar measure.

Typically, the Lie group $G$ is taken to be the dynamical Lie group, which can be found using the corresponding dynamical Lie algebra $\mathfrak{g}$ and the Lie exponential map: $G = e^{\mathfrak{g}}$. The dynamical Lie algebra (DLA) captures the ultimate expressivity of the circuit, and is defined as
\begin{equation}
    \mathfrak{g} = \langle iH_1,..., iH_L\rangle_{\mathrm{Lie}},
\end{equation}
where $H_i$ and $L$ are as in Equation \ref{eq:general_pqc}, and $\langle \rangle_{\mathrm{Lie}}$ is the Lie closure.

The 2-design condition makes possible a key translating step in current plateau theory. Let $U(\boldsymbol{\theta})$ be a parametrized quantum circuit of depth $L$, and let $(\mathcal{E}_L, \nu)$ be its associated ensemble of unitaries. If $(\mathcal{E}_L, \nu)$ forms a $G$ 2-design, then we can write the following, where $\mu$ is the Haar measure over $G$:
\begin{equation}
     \mathbb{E}_{\nu}[C(U(\boldsymbol{\theta}))]
     =
     \mathbb{E}_{\mu}[C(U(\boldsymbol{\theta}))]
\end{equation}
\begin{equation}
     \mathbb{E}_{\nu}[C(U(\boldsymbol{\theta}))^2]
     =
     \mathbb{E}_{\mu}[C(U(\boldsymbol{\theta}))^2]
\end{equation}
It follows that we can express $\mathrm{Var}_{\boldsymbol{\theta}}[C(\boldsymbol{\theta})]$ in terms of the Haar integral over $G$. The move from the circuit's measure (typically taken to be the uniform distribution over the product space of parameters) to the Haar measure is a critical step in barren plateau theoretical analysis, as the Haar integral can be further analyzed via Weingarten calculus and Schur's lemma, while evaluating the former is generally intractable \cite{larocca_barren_2025, ragone_lie_2024, fontana_adjoint_2024, diaz_showcasing_2023, anschuetz_unified_2025}.

\subsection{The Dynamical Lie Algebra and Barren Plateau Theory}
\label{sec:bp_theory}

We are now prepared to review some important theoretical results from the barren plateau literature. McClean et al. discovered that any random parametrized quantum circuits satisfying a sufficiently strong 2-design condition will with high probability exhibit barren plateau behavior \cite{mcclean_barren_2018}. Cerezo et al. proved that, given a VQA with an alternating layered ansatz forming a 2-design, a cost function defined with global observables will induce barren plateaus \cite{cerezo_cost_2021}. Ragone et al. and Fontana et al. independently derived closed-form expressions for the variance of the loss function for parametrized quantum circuits with strong 2-design conditions \cite{ragone_lie_2024,fontana_adjoint_2024}. Crucially, both Ragone et al. and Fontana et al. assume a looser $\varepsilon$-approximate 2-designs rather than exact 2-designs and derive a circuit depth at which a PQC will form an $\varepsilon$-approximate 2-design. The authors also assume a Lie algebra supported ansatz (LASA); that is, the observable $iO$ belongs to the dynamical Lie algebra $\mathfrak{g}$ generated by the circuit. These results have since been extended to hold outside the LASA assumptions, though they still retain the 2-design condition \cite{diaz_showcasing_2023, anschuetz_unified_2025}. 

We focus on Ragone et al. \cite{ragone_lie_2024}, as their closed form expression for the loss landscape variance has been highly influential in current 2-design theoretical approaches to predicting barren plateaus. For a PQC of the form given in Equation \ref{eq:general_pqc}, with initial state $\rho$ and observable $O$, the variance of the cost function $C(\boldsymbol{\theta})$ may be expressed as:
\begin{equation} \label{eq:ragone}
    \mathrm{Var}_{\boldsymbol{\theta}}[C(\boldsymbol{\theta})]
    =
    \sum_{j=1}^{k-1}\frac{
    \mathcal{P}_{\mathfrak{g}_j}(\rho)\mathcal{P}_{\mathfrak{g}_j}(O)
    }
    {
    \dim(\mathfrak{g}_j)
    }
\end{equation}
where $\mathfrak{g}$ is the dynamical Lie algebra induced by the PQC, $\mathfrak{g}_j\subseteq \mathfrak{g}$ are simple ideals of $\mathfrak{g}$, and $\mathcal{P}_{\mathfrak{g}_j}(\rho)$,  $\mathcal{P}_{\mathfrak{g}_j}(O)$ are the $\mathfrak{g}_j$-purities of $\rho$ and $O$, respectively (see Appendix \ref{app:bp-theory} for a definition of $\mathfrak{g}$-purity).

Importantly, as Equation \ref{eq:ragone} indicates, if the dimension $\mathrm{dim}(\mathfrak{g})$ increases exponentially with respect to the number of qubits, we have a barren plateau (see \cite{ragone_lie_2024} Corollary 1). Unfortunately, exponential dimension DLAs are very common, occurring for all but an exponentially vanishing fraction of problem instances for QAOA applied to the Max-Cut problem \cite{mao2025qaoamaxcutbarrenplateausgraphs}. We conjecture that the DLA behaves similarly for QAOA MIS, based both on the rarity of lower-dimension DLAs \cite{allcock_dynamical_2024, wiersema_classification_2023}, and our own preliminary empirical analysis (see Appendix \ref{app:dla-dimension-empirics}). In conjunction with Ragone et al., this result means that QAOA Max-Cut (and, we conjecture, MIS) encounters exponentially large DLAs for virtually all graphs. Therefore, according to DLA theory, we would almost always observe a barren plateau in QAOA on MIS---provided, of course, that the circuit is deep enough to form a unitary 2-design.

In our analysis, we attempt to problematize this 2-design condition. First, here, we argue along theoretical lines that the 2-design condition is hard to verify in practice, and second, we argue empirically that our results do not conform with the 2-design predictions. Determining whether or not a given PQC forms an ($\varepsilon$-approximate) 2-design is not generally a tractable problem. Direct verification is, of course, not generally viable, as it requires Haar integration over the relevant Lie group (see Equation \ref{eq:2-design}). Ragone et al. and Fontana et al. both proved that sufficiently deep PQCs form $\varepsilon$-approximate $2$-designs \cite{ragone_lie_2024, fontana_adjoint_2024, larocca_diagnosing_2022}. Unfortunately, computing this bound is generally a costly endeavor, as it includes computing an operator norm of a term that likewise requires Haar integration over an ambient Lie group (see Appendix \ref{app:bp-theory}). Finally, the assumption of an $\varepsilon$-approximate 2-design naturally introduces an error term between the real and predicted loss landscape variance. We argue that this error term is often non-trivial, indicating a source for the divergence between the Lie-theoretic asymptotic predictions afforded by the theory and the reality of shallow circuit loss landscape scaling (see Appendix \ref{app:bp-theory}, also see Margeta-Cacace \cite{margeta-cacace2026krylovlie}). Crucially, even granting that sufficiently deep circuits 
form $\varepsilon$-approximate 2-designs, the required depth grows with $N$ for any fixed $\varepsilon$ (see Eq.~\ref{eq:ragone-bound}), meaning that constant-depth ans\"{a}tze become progressively worse approximations to 
the 2-design condition as the system scales. The unworkability of 2-design theoretic barren plateau theory in shallow circuits hence motivates our empirical approach of study.

\subsection{Related Work}

Given the significant barrier barren plateaus pose to quantum advantage, there naturally exists recent research analyzing the presence of barren plateaus through numerical simulations.
While some results observe a barren plateau landscape, more recent literature suggests that alternative behaviors exist, particularly for shallow circuits. In some shallow, under-parameterized quantum circuits, the loss landscape of VQAs are ``bumpy" and trap-laden, making it challenging to reach global minima \cite{anschuetz_quantum_2022}. Our results align compatibly with these findings by identifying an increase in gradient variance as the system size scales for the QAOA-based MIS solver applied to shallow circuits.

In shallow circuit depths, Volkoff and Coles, proved that using correlated parameters or local cost functions with QAOA-like ansätze can avoid barren plateaus \cite{volkoff_large_2021}. Most interestingly for the purposes of this paper, they construct simple quantum circuits that display \textit{increasing} variance with respect to the number of qubits on shallow circuit depths ($L=4$), albeit increasing $\mathcal{O}(\sqrt{N})$ instead of $\mathcal{O}(\mathrm{poly}(N))$. These results offer a theoretical precedent for the non-DLA empirics presented in our paper.

There has been recent research on the ability of EHMs in predicting the performance of VQAs on MIS. Prompting our use of a graph convolutional neural network as an EHM, Sohrabizadeh et al. built a graph neural network (GNN) to predict the performance of the Rydberg atom array (RAA) quantum computer on an MIS instance through two metrics of hardness: the probability of the algorithm finding a maximum independent set and the approximation ratio of the independent set generated by the algorithm \cite{sohrabizadeh_gnn-based_2024}. (The approximation ratio is defined as the observed independent set size generated by the algorithm divided by the size of a truly maximum independent set.) Both of these quantities are difficult to compute exactly and are therefore of interest to estimate.

While this work showed the GNN was effective in predicting metrics associated to the \textit{performance} of the RAA-based MIS optimizer, we particularly focus on the ability of machine learning models to predict \textit{barren plateau} behavior in VQA-based MIS optimizers by looking at metrics of hardness directly tied to the gradient variance. This approach allows us to predict whether an MIS instance will induce a barren plateau, or a high-variance ``cragged terrain" landscape. We additionally examine the performance of random forest regression and feedforward neural networks, which use explicit graph features for prediction in contrast to the graph topology used by GNNs. Our methodology therefore permits deeper analysis of the relationship between classical graph theory and quantum trainability, especially through the use of Gini feature importances of our random forest model.

\section{Methods}
\label{sec:methods}

\subsection{Graph Generation}
\label{subsec:graph-generation}

We included all graphs of order up to 7, and randomly sampled $\sim$3,600 graphs of orders 8-20 uniformly from 10 different random graph distributions, tested on circuit depths of $p \in \{1, 3, 5, 10\}$. Ultimately, this amounted to 4,900 distinct graphs and 19,600 problem instances. These distributions were comprised of the Erd\H{o}s-R\'{e}nyi, Watts-Strogatz, and Barab\'{a}si-Albert models with varying parameter choices (details may be found in Appendix \ref{app:graph-details}), in order to generate a diverse array of problem instances. The Erd\H{o}s-R\'{e}nyi random graph model has been used extensively to benchmark the performance of QAOA (e.g., \cite{crooks_2018, appliedmath6020024, Weggemans2022solvingcorrelation}). The Watts-Strogatz and Barab\'{a}si-Albert models are included to collect data on graphs with behavior not typically exhibited by Erd\H{o}s-R\'{e}nyi graphs, namely high clustering and scale-free degree distributions, respectively. These latter two models have also appeared in the literature \cite{bhat_benchmarking_2026} in the context of QAOA, but not as frequently as the Erd\H{o}s-R\'{e}nyi model.

In addition to a dataset of random graphs, we studied a structured dataset consisting of all 2245 vertex-transitive graphs up to order 20. For these, due to an observed significant rise in the difficulty of classically simulating QAOA at larger depths for vertex-transitive graphs, we tested on circuit depths of $p \in \{ 1, 2, 3, 4\}$, yielding a total of 8,980 problem instances \footnote{Aggregating with the randomly generated graph dataset and accounting for overlap, we had in all $\sim$23,000 problem instances.}. A vertex-transitive graph is a graph whose automorphism group acts transitively on the collection of vertices. In other words, any vertex can be mapped to any other vertex by a symmetry of the graph. Vertex-transitive graphs thus have a highly homogeneous structure: in particular, they form a subclass of the class of regular graphs. The dataset of vertex-transitive graphs was taken from the Encyclopedia of Graphs \cite{encyclopedia_of_graphs}, an online database. 

\subsection{Graph Features}
\label{sec:graph-features}
The graph features we computed are listed in Table \ref{tab:features}. The output of each numerical feature is a single real number. The features listed under the statistical type are set-valued: e.g., each vertex of a graph has its own degree, and the Laplacian spectrum consists of a set of eigenvalues. For each of the statistical features, the mean, standard deviation, skewness, minimum, and maximum of the corresponding feature were calculated. The binary features are $0,1$ valued and signify whether a graph has a certain property. 
\begin{table*}[t] 
\caption{\label{tab:features}Features computed for each graph. Bold features are integer-valued; unbolded features are continuous.}
\begin{ruledtabular}
\begin{tabular}{ll}
\text{Type} & \text{Feature}  \tabularnewline
\hline
\addlinespace[1ex]
Numerical & \begin{tabular}[t]{@{}l@{}} 
\textbf{Number of vertices}, \textbf{number of edges}, edge density, \textbf{diameter}, \textbf{radius}, \textbf{vertex connectivity}, \\
\textbf{edge connectivity}, global clustering coefficient, \textbf{treewidth}, average path length, \textbf{circuit rank}, \textbf{girth}, \\
algebraic connectivity, von Neumann entropy, harmonic diameter, Haemers bound. \end{tabular}  \tabularnewline

Statistical & \begin{tabular}[t]{@{}l@{}} \textbf{Degree}, local clustering coefficient, betweenness centrality, harmonic centrality, adjacency spectrum, \\
Laplacian spectrum, \textbf{core number}. \end{tabular}    \tabularnewline

Binary & \textbf{Planar}, \textbf{chordal}, \textbf{claw-free}.     \tabularnewline
\end{tabular}
\end{ruledtabular}
\end{table*}

The treewidth was computed approximately using the Min Fill heuristic from the NetworkX library \cite{networkx}. The impact of treewidth on the runtime of solving MIS for unit disk graphs on neutral atom quantum processors using adiabatic quantum computing has been empirically studied in the past \cite{cazals_treewidth2025}. The authors there found that runtime increased as treewidth increased until very high values of treewidth, whereafter runtime decreased, justifying its use as a graph feature.

As for other selected features, for disconnected graphs, we defined the diameter and radius to be the sum of the diameters and radii respectively of the connected components. Meanwhile, the Haemers bound \cite{haemers_1979} is an upper bound on the independence number of a graph, or the cardinality of its maximum independent set.

\subsection{QAOA Implementation}
\label{sec:qaoa-impl}

We implemented JPMorgan Chase's open-sourced quantum optimization toolkit QOKit \cite{QOKit} to classically simulate QAOA for the SciPy-native Broyden–Fletcher–Goldfarb–Shanno (BFGS) optimizer with a max iteration count of 1000 \footnote{Preliminary testing was done on other optimizers such as Adam \cite{adam} and the modified conjugate natural gradient method (CQNG) \cite{cqng}, but BFGS seemed to be more performant than these alternatives.}. We chose to forgo considering the effects of finite, repeated, measurement to guarantee that our classification of VQA landscapes was uncompromised by sampling variance. This also helped limit the cost of classically simulating QAOA at the scale of thousands of instances, particularly for problems that have barren plateaus. The MIS Hamiltonian was constructed according to Equation \ref{eq:mis-hamiltonian}, and the Lagrange multiplier was set to $\lambda=1$.

We employ two primary hardness metrics:
\begin{enumerate}
  \item $N_\mathrm{calls}$: the total number of calls to the (classically simulated) QAOA
        circuit during optimization, computed as $\sum_\mathrm{iterations} N_\mathrm{evals/iter}$.
        This metric is hardware-agnostic, eliminates compiler-specific variation, and
        correlates directly with cost on real quantum hardware.
  \item Mean gradient variance (here calculated along the training path), defined as
\begin{equation}
\overline{\mathrm{Var}}_{\boldsymbol{\theta}}[\partial_\mu C(\boldsymbol{\theta})]
  = \frac{1}{N_\theta} \sum_{i=1}^{N_\theta}
    \mathrm{Var}\!\left[\frac{\partial C}{\partial \theta_i}\right],
  \label{eq:mean-grad-var-def}
\end{equation}
\end{enumerate}

The mean variance of the loss landscape gradients serves as our primary barren plateau diagnostic.
By construction, whenever all individual gradient variances vanish
exponentially in the number of qubits, i.e.,
$\mathrm{Var}[\partial_{\mu} C] \in \mathcal{O}(1/b^{N})$ for some $b>1$ for every
$\partial_\mu$, the mean gradient variance \eqref{eq:mean-grad-var-def}
is also exponentially small.
Thus any barren plateau detected via the conventional per-parameter
variance or cost variance criteria is also detected by
$\overline{\mathrm{Var}}_{\boldsymbol{\theta}}[\partial_\mu C(\boldsymbol{\theta})]$.
Conversely, $\overline{\mathrm{Var}}_{\boldsymbol{\theta}}[\partial_\mu C(\boldsymbol{\theta})]$ can reveal ``partial''
barren plateaus, in which a non-negligible subset of directions already
exhibits exponentially small gradients, even if global cost fluctuations
have not yet fully concentrated. Moreover, BFGS gradient magnitudes decrease superlinearly per step under standard assumptions (strong convexity, Lipschitz Hessian, Wolfe line search; see Nocedal and Wright's chapter on quasi-Newton methods \cite{NocedalWright2006}), meaning that variances computed along the training path are expected to be smaller than those obtained from sampling over the entire domain.\footnote{We use this only as a heuristic; nonconvex counterexamples certainly exist, but they do not undermine the qualitative point that small mean gradient variance along the training path is a conservative indicator of flatness.} 

Since throughout we are interested in any behavior that can be bounded above by barren plateau-like decay in our training directions (any decay faster than a barren plateau is clearly at least as damaging to VQA training feasibility as a barren plateau), the various indicators (per-parameter gradient variance, cost variance, and mean gradient
variance) are equivalent at the level of asymptotic behaviour, differing
only by constant prefactors that are asymptotically irrelevant. Moreover, using $\overline{\mathrm{Var}}_{\boldsymbol{\theta}}[\partial_\mu C(\boldsymbol{\theta})]$ along the training path has the advantage of having all its samples collected automatically during training, limiting the computational expense of classical simulation, as well as being easier to interpret and analyze than each gradient component taken individually.

\subsection{Identifying Barren Plateaus} 
\label{sec:identifyBP}
Because the MIS problem changes fundamentally with $N$ (the number of qubits, and, in this case, the number of vertices), we cannot directly compare the
same instance at different system sizes. Instead, we use graph feature clustering
as a proxy for fixing the problem structure.  For each graph feature $F$ (Section \ref{sec:graph-features}) and each circuit
depth $p$, we partition instances into groups of similar feature values and examine the
scaling of $\overline{\mathrm{Var}}_{\boldsymbol{\theta}}[\partial_\mu C(\boldsymbol{\theta})]$ against $N$ within each group.

Continuous features were clustered using HDBSCAN as implemented by the Python package Scikit-learn \cite{hdbscan} to avoid
artificial truncation at cluster boundaries \footnote{For the randomly-generated dataset, the minimum cluster size for HDBSCAN was taken to be the integer cast of $1/60$ times the dataset size for the current $p$, and the minimum number of samples was the integer cast of one half of the minimum cluster size. Similarly, for the vertex-transitive dataset, the minimum cluster size was $1/30$ times the dataset size and the minimum sample number was one half of that quantity. These choices of hyperparameters seemed to provide a good balance between merging nearby clusters with similar densities and not overgrouping sets that were too unrelated.}; discrete features were grouped into sets where their integer value was identical.  Clusters with fewer than four distinct values of $N$ were discarded.  For each
cluster, we computed an exponential regression of $N$ against $\overline{\mathrm{Var}}_{\boldsymbol{\theta}}[\partial_\mu C(\boldsymbol{\theta})]$
and recorded the Pearson correlation coefficient $r$ associated with the slope of the linear equation of the logarithm of the best fit exponential. Clusters with $r$ closer to $-1$ then better matched $\overline{\mathrm{Var}}_{\boldsymbol{\theta}}[\partial_\mu C(\boldsymbol{\theta})]$ asymptotics representing barren plateaus, whereas clusters closer to $+1$ exhibited asymptotics more indicative of cragged terrains. However, it is important to recognize that these empirical correlations do not represent exact asymptotics, and that caution must therefore be observed in extrapolating their indicated behavior. This is of utmost importance in the interpretation of the cragged terrains, as Popoviciu's inequality coupled with trivial upper and lower bounds for the MIS expectation value ($N^2$ and $-N$) immediately shows that the variance of the problem landscape can grow no faster than $O(N^4)$ in the extreme case. Nonetheless, this does not contradict the utility of the Pearson $r$, since a growing polynomial can be well-approximated by an exponential of appropriate base for sufficiently small $N$.

\subsection{Empirical Hardness Models}

After extracting our set of fifty-two tabular graph features from our graph data as described in Section \ref{sec:graph-features}, we can train a machine learning model to predict the mean gradient variance from graph problem instances. Since we are learning and modeling a non-linear problem, we train on three types of non-linear models: random forest regression, a feedforward neural network, and a graph convolutional neural network. We discuss our rationale for choosing these three models as well as their architecture.

\subsubsection{Random Forest}

Random forest regression is a supervised ensemble learning method that creates several decision trees, each trained on a random subset of data using a random subset of features. Each of these trees makes their own prediction. The final prediction is the average of each the prediction from each decision tree. We choose random forest regression for their simplicity and their prior effective use as empirical hardness models \cite{leyton-brown_empirical_2009, leyton-brown_understanding_2014}. Critically, however, random forest regression is typically weaker at generalizing to output results not present in the training data, since the prediction is the simply of the average of several decision trees \cite{biau2015randomforestguidedtour}. This severely limits the ability of the random forest regression to generalize well to higher qubit problems.

\subsubsection{Feed Forward Neural Network}

We use a feed forward neural network as a baseline neural network model. We use three hidden layers with $32$, $32$, and $16$ neurons, respectively, each followed by the ReLU activation function. We found that increasing the number of layers and nodes had a minimal impact on the performance of the model. Based on results from Optuna \cite{akiba_optuna_2019}, we found the Adam optimization algorithm \cite{adam} with a learning rate of $\eta = 0.006$ to be most effective.

\subsubsection{Graph Convolutional Neural Network}

The empirical hardness models implemented thus far have been trained on a set of fifty-two tabular graph features. Naturally, one may worry that our choice of graph features does not completely capture the graph structure. One solution to this problem comes in the form of a graph neural network-based empirical hardness model, which takes in the graph structure (vertices and edges) directly as an input.

Since the model also needs to access non-graph structure data, such as the circuit depth $p$, we insert these values as node features of the input graph. We found that including salient tabular graph features improved model performance, so these were also inserted as node features.

We implemented three GraphSAGE \cite{hamilton2017inductive} layers, the first with top-K pooling and the last two with the ReLU activation function. This was followed by two linear layers, along with mean global pooling and dropout layers. Testing with different graph convolutional aggregation methods (i.e., GANs, Graph Attention, Graph Isomorphism) anecdotally yielded slightly worse results, and changing the number of convolutional and linear layers offered zero to negative improvement. We found the current implementation to be a straightforward and powerful solution, though we leave open the possibility of a better architecture.

\section{Results}
\label{sec:results}

\subsection{Empirical Hardness Models}
For each of the three empirical hardness models, we performed 10 trials and randomly divided our data for each trial into training and testing data with an $80:20$ split.

\begin{table}[htbp]
  \caption{EHM test $R^2$ scores (mean $\pm$ std over 10 trials, 80/20 split).}
  \label{tab:ehm-perf}
  \begin{ruledtabular}
    \begin{tabular}{lcc}
      Model & $\overline{\mathrm{Var}}_{\boldsymbol{\theta}}[\partial_\mu C(\boldsymbol{\theta})]$ & $N_\mathrm{calls}$ \\ \hline
      \addlinespace[1ex]
      Random Forest & $0.973\pm0.010$ & $0.840\pm0.016$ \\
      Neural Network & $0.837\pm0.025$ & $0.823\pm0.011$ \\
      GNN & $0.959\pm0.009$ & $0.838\pm0.017$ \\
    \end{tabular}
  \end{ruledtabular}
\end{table}

\textbf{Comparison of Models:} The predicted versus actual target values for the most accurate run of each model are plotted in Figure \ref{Comparison:calls} along with the mean $R^2$ score across 10 trials. We found the random forest regression model to be most effective at modeling this metric of hardness with an average $R^2$ of $0.840$. Similar results were observed when this process was repeated to evaluate the performance of the models on predicting the mean variance of the gradients. The predicted versus actual target values of each model are plotted in Figure \ref{Comparison: mvg} along with the mean $R^2$ score across 10 trials (each plot represents the performance of a randomly chosen trial). Possessing an average $R^2$ of $0.973$, the random forest regression model continued to outperform other model classes when trained on this metric of hardness, although the GCNN was not a significantly worse predictor. Analogous to the results of \cite{leyton-brown_understanding_2014}, our findings demonstrate the effectiveness of random forest regression-based empirical hardness models in predicting both metrics of hardness. This aligns with past work demonstrating that more classical machine learning methods like random forests can sometimes exhibit much more reliable and easily-tuned performance over GNNs \cite{Ting_2021, bechlerspeicher2024graphneuralnetworksuse}.

\begin{figure}[htbp]
\centering    
        \includegraphics[width=0.9\columnwidth]{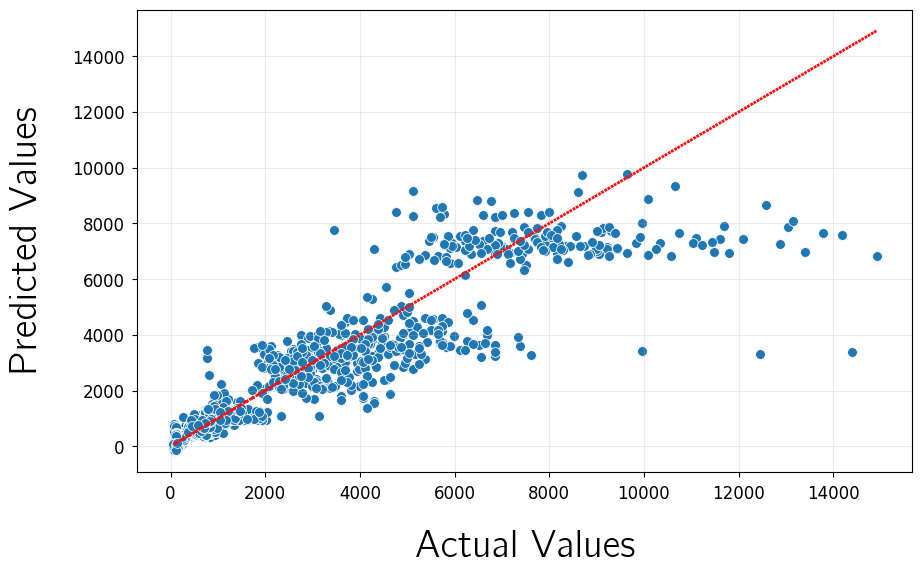}

\caption{Performance of the graph convolutional neural network when training on N from 1 to 16 and predicting the number of calls to the quantum circuit on N from 19 to 20. The stratification into three general sections is possibly due to the presence of three distinct graph distributions in our data and the model categorizing its predictions to a significant degree based off this tricategorization. }\label{Generalization}
\end{figure}

To test our models' ability to extrapolate to higher qubit counts, we used a training dataset of randomly generated graphs with $N$ ranging from 1 to 16 and a test dataset with $N$ ranging from 19 to 20. We found that all three models performed comparably to non-generalization tasks, albeit with an unusual trifurcated plot (see Figure \ref{Generalization}).\footnote{The random forest regression model had an $R^2$ value of $0.878\pm 0.002$, the neural network had an $R^2$ value of $0.794\pm 0.036$, and the graph neural network had an $R^2$ value of $0.851 \pm 0.006$.} However, when replicating these experiments for the mean gradient variance as the desired output variable, we observed poor prediction accuracy across the board.\footnote{The random forest regression model had an $R^2 = 0.621\pm 0.004$, the neural network had an $R^2$ value of $0.647\pm  0.062$, and the graph neural network had an $R^2$ value of $0.641 \pm 0.071$.} It may be the case that our choice of features was relatively ill-suited to the prediction of the mean variance of the gradients while less so for the number of quantum calls, leading to the models overfitting when learning the mean variance of the gradients. To address this issue in the future, one may wish to use features that more explicitly depend on the algebraic structure of the VQA (especially some compressed variant of the commutator structure of the generators).


\begin{table}[htbp]
  \caption{Gini feature importances (top features) for the random forest EHM predicting
  $N_\mathrm{calls}$ and $\overline{\mathrm{Var}}_{\boldsymbol{\theta}}[\partial_\mu C(\boldsymbol{\theta})]$.  Circuit depth $p$ dominates
  both targets.}
  \label{tab:feature-importance}
  \begin{ruledtabular}
    \begin{tabular}{lcc}
      Feature & $N_\mathrm{calls}$ & $\overline{\mathrm{Var}}_{\boldsymbol{\theta}}[\partial_\mu C(\boldsymbol{\theta})]$ \\ \hline
      \addlinespace[1ex]
      Circuit depth $p$ & $0.552$ & $0.505$ \\
      Laplacian spectrum skewness & $0.088$ & $0.001$ \\
      Adjacency spectrum skewness & $0.065$ & $0.006$ \\
      Mean core number & $0.060$ & $0.000$ \\
      Algebraic connectivity & $0.030$ & $0.003$ \\
      Laplacian spectrum mean & $0.016$ & $0.043$ \\
      Adjacency spectrum max & $0.014$ & $0.073$ \\
      Adjacency spectrum std dev & $0.014$ & $0.060$ \\
      Circuit rank & $0.002$ & $0.072$ \\
      Mean harmonic centrality & $0.002$ & $0.071$ \\
    \end{tabular}
  \end{ruledtabular}
\end{table}

\subsection{Barren Plateaus \& Cragged Terrains}
\subsubsection{Cragged Terrain Prevalence}

\begin{figure}[h!]
    \centering
    \subfloat[Randomly Generated Graphs]{
    \includegraphics[width=0.48\textwidth]{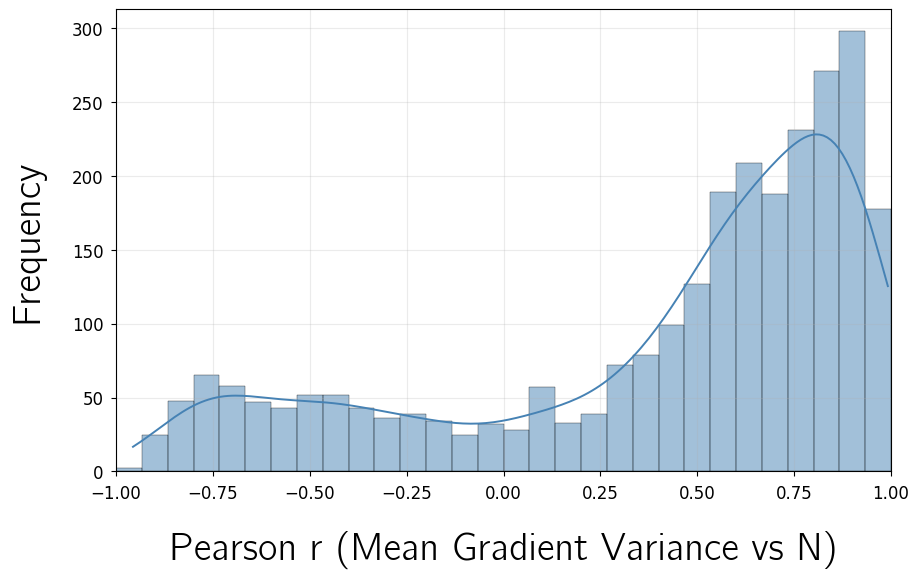}}
    \label{fig:er_histogram}
    \subfloat[Aggregated Data]{
    \includegraphics[width=0.48\textwidth]{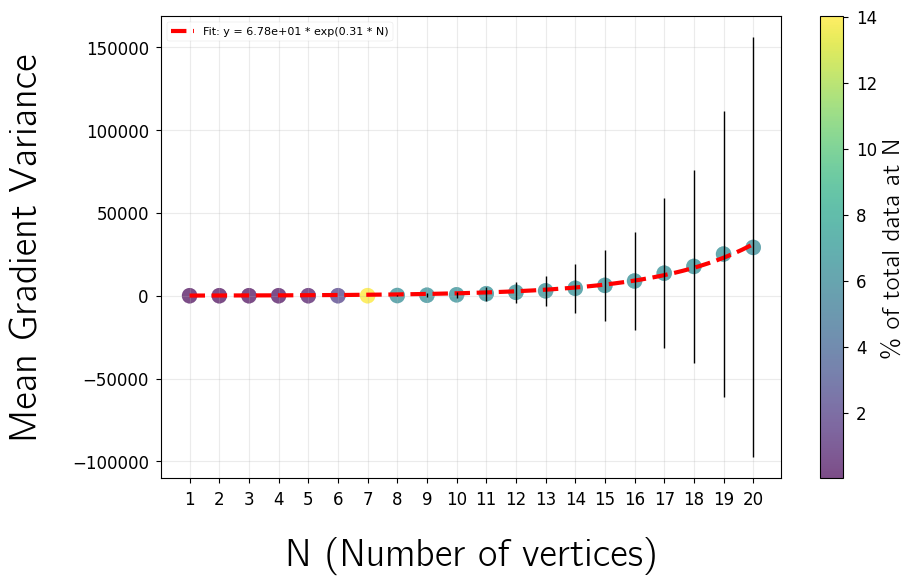}}
    \label{fig:er_aggregate}
    \caption{(a) \textbf{Pearson $\mathbf{r}$ Distribution.} Pearson correlation coefficients of the mean variance of gradients against $N$ for each cluster of random graphs. The highly left-skewed distribution indicates widespread ``cragged terrain'' behavior. (b) \textbf{Aggregated Data is a Cragged Terrain.} Plotting the mean variance of the gradients for each value of $N$ (each point and its error bars represent the average value and standard deviation of the subset of the data at that value of $N$) for $p \in \{1,3,5,10\}$, we observe a positive exponential relationship with Pearson correlation coefficient $r = 0.977$. Curiously, the standard deviations at each $N$ also had a positive exponential correlation, this time with a correlation coefficient of $0.963$.}
\end{figure}

\begin{figure}[h!]
    \centering
    \subfloat[Vertex Transitive Graphs]{
    \includegraphics[width=0.48\textwidth]{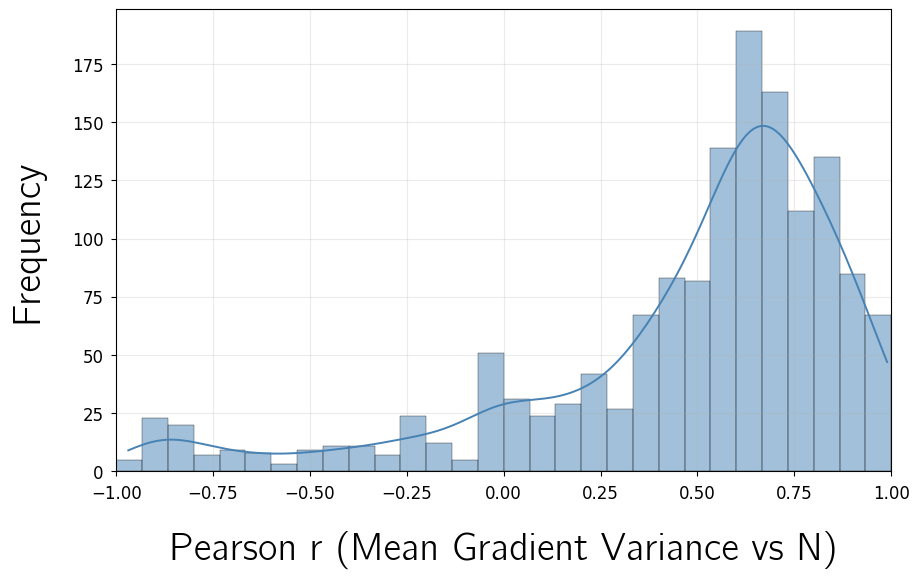}}
    \label{fig:vt_histogram}
    \subfloat[Aggregated Data]{
    \includegraphics[width=0.48\textwidth]{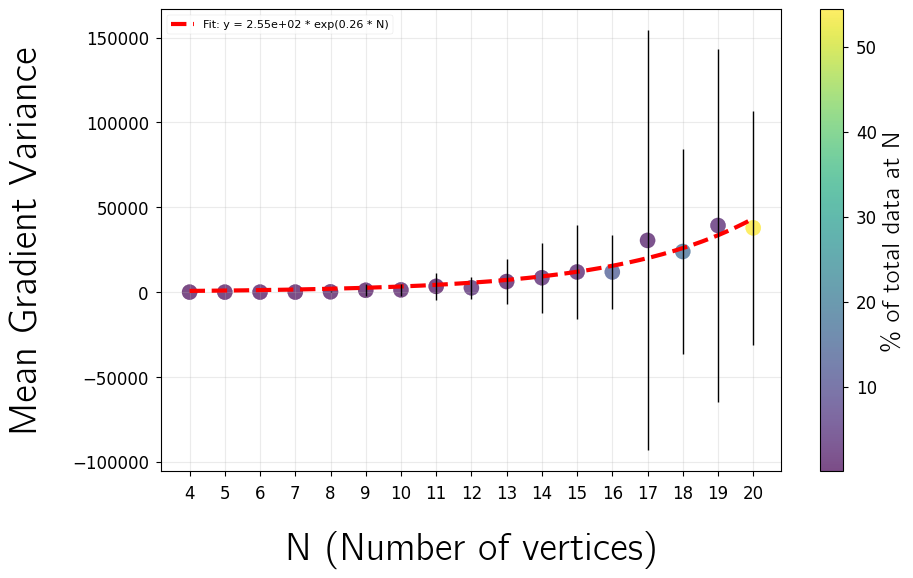}}
    \label{fig:vt_aggregate}
    \caption{(a) \textbf{Pearson $\mathbf{r}$ Distribution for Vertex Transitive Graphs.} Pearson correlation coefficients of the mean variance of gradients against $N$ for each cluster of random graphs. We a similarly left-skewed distribution, indicating that vertex transitive graphs also exhibit widespread cragged terrain behavior. (b) \textbf{Aggregated Vertex Transitive Data also is a Cragged Terrain.} Plotting the mean variance of the gradients for each value of $N$ for $p \in \{1,2,3, 4\}$, we observe a positive exponential relationship with Pearson correlation coefficient $r = 0.941$. Although the error bars do not observe the same exponential relation as observed in the random graphs (see Figure \ref{fig:er_aggregate}), we notice unusual spikes on numbers that have few prime factors. This is notable, as the number of vertex transitive graphs with highly divisible orders tend to be considerably higher \cite{royle_constructing_1989}.}
\end{figure}

Using the clustering procedure of Sec.~\ref{sec:identifyBP}, we observe a striking asymmetry
in the distribution of Pearson correlation coefficients (Fig.~\ref{fig:er_histogram}).  For
randomly generated graphs, only $0.998\%$ of clusters satisfy $r < -0.85$
(indicative of barren plateaus), while ${17.115\%}$ of clusters satisfy $r > 0.85$
(cragged terrains). At a looser threshold ($|r| > 0.75$), the contrast remains striking:
$3.689\%$ are classified as BPs versus $29.392\%$ as cragged terrains.  This left-skewed distribution of Pearson coefficients is robust across both random and vertex-transitive datasets (Fig.~\ref{fig:vt_histogram}). Both aggregate datasets (upon which no clustering has been performed) also exhibited cragged terrain behavior. Recalling now that $\overline{\mathrm{Var}}_{\boldsymbol{\theta}}[\partial_\mu C(\boldsymbol{\theta})]$ is strictly more sensitive to BP-like behavior than either $\mathrm{Var}_{\boldsymbol{\theta}}[C(\boldsymbol{\theta})]$ and ${\mathrm{Var}}_{\boldsymbol{\theta}}[\partial_\mu C(\boldsymbol{\theta})]$ (Section \ref{sec:qaoa-impl}), we see that these these statistics are, if anything, an \emph{overrepresentation} of BP presence in QAOA on MIS.

In broad consideration of these facts, and because the random graphs constitute a more representative subset of graphs up to $N=20$, and moreover the vertex-transitive dataset exhaustively catalogs every graph of that particular, highly-structured type to that same maximum order, we have strong reason to suspect that our results are not mere artifacts of our graph sampling protocol or clustering algorithm. The observed prevalence of cragged terrains---in addition to the apparent dearth of barren plateaus---appears to reflect a genuine structural phenomenon. 

\begin{figure}[h!]
    \centering
    \subfloat[\textbf{A Barren Plateau.} Example of a cluster exhibiting
    barren plateau behavior when clustering on the von Neumann entropy for
    $p=3$. As the number of qubits increases, we observe an exponential
    decrease in the mean variance of the gradients with a Pearson coefficient
    of $-0.922$.]{%
        \includegraphics[width=0.48\textwidth]{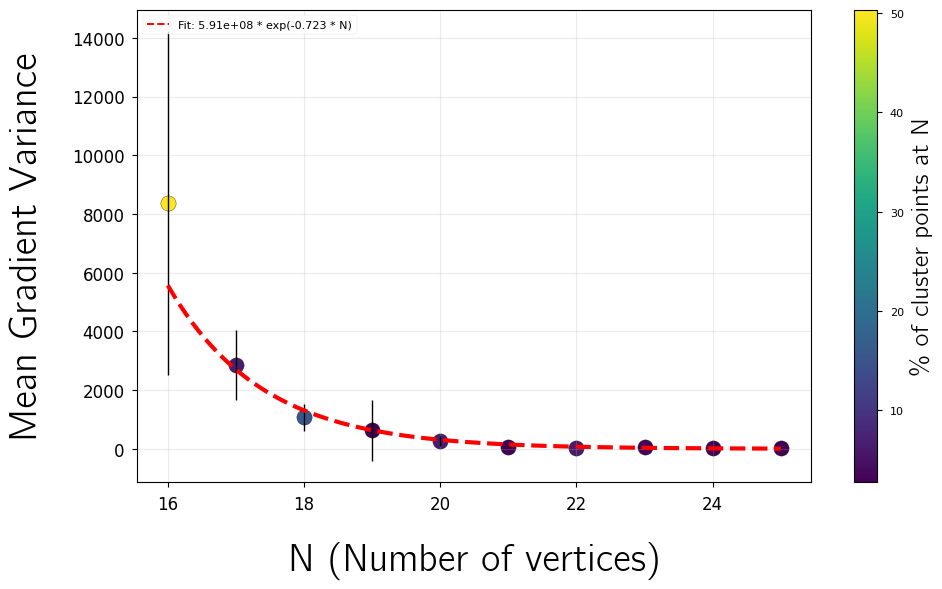}
        \label{fig:bp_example}
    }
    \hfill
    \subfloat[\textbf{A Cragged Terrain.} Example of a cluster exhibiting
    cragged terrain behavior when clustering on edge density for $p=5$. As the
    number of qubits increases, we see an exponential increase in the mean
    variance of the gradients with a Pearson coefficient of $0.992$.]{%
        \includegraphics[width=0.48\textwidth]{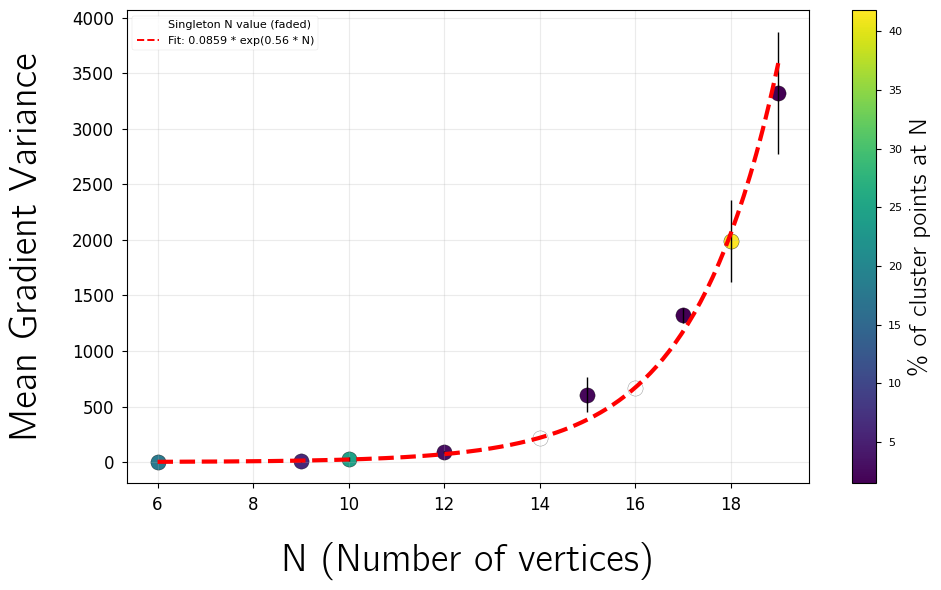}
        \label{fig:ct_example}
    }
    \caption{\textbf{Example of barren plateau and cragged terrain cluster behaviors.}}
    \label{fig:bp_ct_example}
\end{figure}

As mentioned previously, the aggregate data of random graphs clearly manifests a cragged terrain. Plotting the mean of $\overline{\mathrm{Var}}_{\boldsymbol{\theta}}[\partial_\mu C(\boldsymbol{\theta})]$ at each value of $N$ for $p\in\{1,3,5,10\}$ yields a very strong positive exponential correlation (Pearson $r = 0.977$). This phenomenon persists even after splitting the data into groups of different $p$ values. The vertex-transitive dataset mirrors these findings ($r = 0.941$). Intriguingly, the standard deviations at each $N$ for the random dataset also grow exponentially ($r = 0.963$), though this effect is not present in vertex-transitive graphs.\footnote{Instead, there is a highly irregular pattern in the standard deviations across $N$, potentially related to the unique dependence of vertex-transitive graphs on the number of factors of their order.} Still, as explained in Section \ref{sec:identifyBP}, although these curves are fit well by exponential regressions, such fits are statistically indistinguishable from a low-degree (less than or equal to quartic) polynomial increase for $N\leq 20$, and we must conservatively interpret the observed growth accordingly.\footnote{Additionally, the possibility that the variance scaling follows a log-normal distribution as opposed to a power law should not be discounted.} This contrasts with prior analyses that report the opposite trend of barren plateaus in the aggregate
\cite{mao2025qaoamaxcutbarrenplateausgraphs, larocca_diagnosing_2022, yao2026gradientanalysisbarrenplateau, Kashif_2024}, suggesting that the behavior observed here is specific to the
shallow-circuit, MIS-on-general-graphs setting and represents a novel departure from conventional expectations.


\subsubsection{Graph Features Correlated with Baren Plateau / Cragged Terrain Behavior}

Several graph features are strongly predictive of BP and cragged terrain cluster membership (Table~\ref{tab:feature-bp-rt}).  For random graphs, the von Neumann graph entropy is the most strongly BP-associated feature (aggregated Pearson
$r = -0.672$), while edge density and harmonic diameter are the most strongly
cragged-terrain-associated features ($r = +0.938$ and $+0.934$, respectively).\footnote{Surprisingly, despite measuring obviously related quantities and the edge density being incredibly CT-predictive, the edge count is weakly BP-associated for random graphs ($r=-0.2779$). One plausible explanation is that edge density measures the fraction of possible interactions that actually exist, making it an intrinsic structural property of the graph invariant of system size, whereas raw edge count conflates graph structure with system size. As $N$ grows, a graph can accumulate many edges while remaining sparse, so the two quantities diverge in precisely the regime where landscape scaling matters most.} These trends also remain steadfast in the vertex-transitive data, which also sees von Neumann entropy being strongly BP-correlated as well as edge density and harmonic diameter being highly CT-correlated. Similarly, the Haemers bound and average path length were associated with CTs across both graph distributions. Harmonic centrality statistics are consistently BP-associated across both random and vertex-transitive datasets. The appearance of multiple spectral statistics (adjacency spectrum, Laplacian spectrum) in the ranked feature table (Table~\ref{tab:feature-importance}) indicates
that spectral graph theory may provide theoretical insight into these landscape behaviors. The consistency of these trends in both the random graph and vertex-transitive datasets suggests that they are not artifacts of a particular choice of graph distribution. Thus, the idea of looking at local connectivity and complexity to predict BPs and global geometric spread to predict CTs in terms of graph features may bear fruit.

\begin{table}[htbp]
  \caption{Top graph features associated with BP and cragged terrain (CT) cluster membership.
  Aggregated Pearson $r$ of $\ln\overline{\mathrm{Var}}_{\boldsymbol{\theta}}[\partial_\mu C(\boldsymbol{\theta})]$ vs.\ $N$ within clusters
  of each feature.}
  \label{tab:feature-bp-rt}
  \begin{ruledtabular}
   \renewcommand{\arraystretch}{1}
    \begin{tabular}{lc}
      Feature & Aggregated $r$ \\ \hline
      \addlinespace[1ex]
      \multicolumn{2}{c}{\textit{BP-associated (random graphs)}} \\
      von Neumann entropy & $-0.6719$ \\
      Mean harmonic centrality & $-0.3661$ \\
      Max harmonic centrality & $-0.3147$ \\
      Number of edges & $-0.2779$ \\
      Min harmonic centrality & $-0.2204$ \\
      \multicolumn{2}{c}{\textit{BP-associated (vertex-transitive)}}  \\
      Mean harmonic centrality & $-0.6534$ \\
      Max harmonic centrality & $-0.6529$ \\
      Min harmonic centrality & $-0.6375$ \\
      von Neumann entropy & $-0.6095$ \\
      Circuit rank & $-0.4384$ \\
      \multicolumn{2}{c}{\textit{CT-associated (random graphs)}} \\
      Edge density & $+0.9377$ \\
      Harmonic diameter & $+0.9343$ \\
      Haemers bound & $+0.8743$ \\
      Min local clustering coeff.\ & $+0.8678$ \\
      Average path length & $+0.8382$ \\
      \multicolumn{2}{c}{\textit{CT-associated (vertex-transitive)}} \\
      Edge density & $+0.9111$ \\
      Harmonic diameter & $+0.8781$ \\
      Average path length & $+0.8524$ \\
      Haemers bound & $+0.7517$ \\
      Radius / Diameter \ & $+0.7276$ \\
    \end{tabular}
  \end{ruledtabular}
\end{table}

\subsubsection{EHMs Predict Barren Plateau and Cragged Terrain Behavior}
\label{sec:bp-pred}

Although the EHMs are trained on individual instance hardness values, it is natural to examine whether they preserve the exponential $N$-scaling relationships characteristic of barren plateaus and the polynomial growth of cragged terrains (which is proffered in our analysis as exponential growth with a base sufficiently close to $1$).  After running all graphs in each cluster through the feed forward neural network, the GCNN, and the random forest regression, the predictions maintain their exponential scaling signatures. Across all clusters with $r < -0.85$ or $r > 0.95$, the Pearson correlation coefficient of the predicted exponential fit matches the true coefficient with a mean percent error of $2.777\%$, ${2.406\%}$, and $2.208\%$, respectively.  This demonstrates that our models have implicitly learned the structural correlates of the expected $N$-scaling, and that EHMs could serve as a computationally inexpensive diagnostic of barren plateaus and cragged terrains. 

\begin{figure}[h!]
    \centering
    \subfloat[Barren Plateau Cluster Prediction]{
    \includegraphics[width=.48\textwidth]{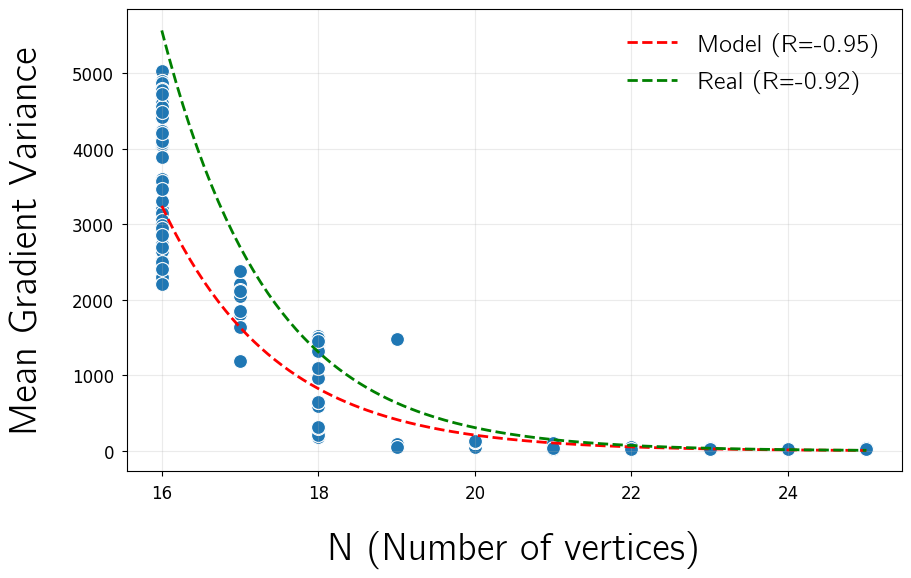}}
    \label{fig:GNNBarrenPlateausA}
    \subfloat[Cragged Terrain Cluster Prediction]{
    \includegraphics[width=.48\textwidth]{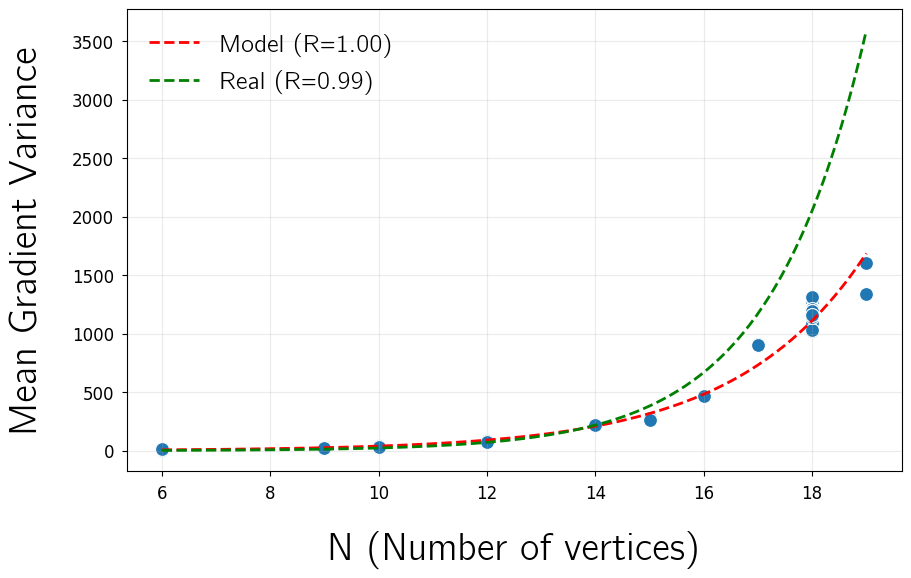}}
    \label{fig:BarrenPlateauGNNB}
    \caption{\textbf{GCNN Predicts Barren Plateaus and Cragged Terrains.} Example of the graph neural network predictions for a cluster of graphs exhibiting a barren plateau and cragged terrain. The clusters are the same as in Figures \ref{fig:bp_example} and \ref{fig:ct_example}: von Neumann entropy for $p=3$ and edge density for $p=5$, respectively. The red fit line is the fit for the model predictions, the green fit line is the original data fit, and the blue data points are predicted values.}
    \label{fig:BarrenPlateauGNN}
\end{figure}

\section{Discussion}
\label{sec:discussion}

Our results indicate a pronounced asymmetry between barren-plateau-like and cragged terrain behavior in shallow-circuit QAOA for MIS. Within the range of graph families examined here, barren-plateau-like clusters are rare, whereas clusters exhibiting increasing variance with system size are comparatively common. Because the data extends only to modest system sizes, we reiterate that we are forced to interpret these trends as finite-size evidence for distinct landscape classes rather than as definitive asymptotic statements. However, given that our vertex-transitive graph dataset was exhaustive up to order $N=20$ and that our random graphs were drawn from many different distributions detailed in Subsection \ref{subsec:graph-generation}, it appears fairly unlikely that these findings will completely disappear with higher qubit counts, though we stress that this scaling could still be indicative of a log-normal distribution instead of genuine power-law behavior since our data only goes to $N=20$. 

In either case, our data provides strong evidence that DLA and design-centric theories of VQA landscapes cannot simply be extended into the shallow circuit regime: despite exponentially large DLAs \ref{app:dla-dimension-empirics}, we observe variance scaling that is diametrically opposed to the presence of barren plateaus. In the cases where DLAs are not exponential (such as in the vertex-transitive graphs), we still continue to see these polynomially growing variances of qualitatively the same form in our landscapes. This calls into question other supposed barren plateau results that stem from DLA theory \cite{mao2025qaoamaxcutbarrenplateausgraphs, larocca_diagnosing_2022}, as without at least significant constraints on the depth and measure of the circuit, an exponential DLA alone cannot be enough to explain the presence or absence of barren plateaus. Even if this scaling result is somehow optimizer-specific, the fact that BFGS is consistently able to avoid barren plateaus often enough along its training trajectories to almost universally observe polynomially increasing variances over 23,000 problem instances implies that there are major aspects of the VQA landscape story being left out by design-centric analysis, and in particular DLA theory.

The empirical hardness models considered here are best viewed with some caution. Although they perform reasonably well when predicting the hardness metrics on in-distribution data, their generalization to larger system sizes, larger circuit depths, and different graph ensembles is poor. As a consequence, their practical utility as general-purpose predictors of instance-wise hardness appears limited.

Their strongest performance instead arises at the level of finite-size loss-landscape asymptotics. While trained only on individual hardness values, the models recover the scaling behavior of clustered graph families with high fidelity over the full range of qubit counts studied here. This suggests that empirical hardness models may be substantially more useful as diagnostics of landscape-scaling regimes than as predictors of raw hardness metrics.

The graph-feature analysis further suggests that these landscape classes are strongly organized by coarse structural properties of the underlying graphs. The repeated appearance of spectral and transport-related quantities among the most informative features points to spectral graph theory as a natural framework for explaining why certain graph families fall into flattening regimes while others exhibit increasingly rough landscapes.

A further statistical feature of the data is the apparent power-law relationship
between the variance of $\overline{\mathrm{Var}}_{\boldsymbol{\theta}}[\partial_\mu C(\boldsymbol{\theta})]$ and its cross-instance
mean, as both are indexed by $N$ within the observed range $N \leq 20$. Since
$\overline{\mathrm{Var}}_{\boldsymbol{\theta}}[\partial_\mu C(\boldsymbol{\theta})]$ is strictly nonnegative, a Tweedie-type variance
law provides a natural statistical description, with the gamma distribution as a
natural benchmark if the fitted variance power is close to two. At minimum, this
indicates that the observable is strongly heteroscedastic and is therefore not
well-described by models that assume variance independent of
$\overline{\mathrm{Var}}_{\boldsymbol{\theta}}[\partial_\mu C(\boldsymbol{\theta})]$ across $N$.

Several directions for future work follow naturally from these observations. The most immediate priority is to validate the finite-size asymptotic results at substantially larger qubit counts. It will also be important to move beyond optimizer-dependent training-path statistics by sampling gradient information more uniformly across the full loss landscape. These steps will allow for more robust verification of our results and ensure that they are not only induced by training dynamics, but are features of the entire loss landscape. Relatedly, it may prove worthwhile to find the necessary conditions for cragged terrains to be trap-laden, as Anschuetz and Kiani have showed that VQAs are swamped with poor local minima (anecdotally, we did observe large variances and relatively suboptimal performance in the approximation ratios of constant-depth QAOA, but do not have sufficient evidence to lay substantive claim to presence of the phenomenon without additional data or analysis) \cite{anschuetz_quantum_2022}. Furthermore, developing models trained directly on asymptotic targets such as scaling exponents or regime labels, perhaps through random-forest regressions or tensor-network methods, may provide a practical route to extending these conclusions to larger systems while clarifying the connection between graph structure and landscape behavior.
\medskip
\begin{acknowledgments}
The authors thank NASA QuAIL and the Institute for Pure and Applied Mathematics (IPAM)
for support through the Research in Industrial Projects for Students (RIPS) program.
This work was supported by NASA and NSF Grant DMS 1925919.  The authors are grateful to
Dr.\ Lucas T. Braydwood (NASA QuAIL) and Dr.\ Aaron Lott (Hewlett Packard Enterprise; formerly NASA QuAIL) for their mentorship, and to Aman Mehta
(UCLA) for academic guidance.
\end{acknowledgments}

\bibliography{references}

\appendix

\section{Barren Plateau Theory: Limitations in the Shallow-Circuit Regime}
\label{app:bp-theory}

\subsection{Limitations of the Ragone et al.\ Framework}

The main result of Ragone et al.\ \cite{ragone_lie_2024} (Theorem 1 therein) expresses
$\mathrm{Var}_{\bm\theta}[C(\boldsymbol{\theta})]$, initial state $\rho$ and observable $O$, exactly:
\begin{equation} \label{eq:ragone-app}
    \mathrm{Var}_{\boldsymbol{\theta}}[C(\boldsymbol{\theta})]
    =
    \sum_{j=1}^{k-1}\frac{
    \mathcal{P}_{\mathfrak{g}_j}(\rho)\mathcal{P}_{\mathfrak{g}_j}(O)
    }
    {
    \dim(\mathfrak{g}_j)
    }
\end{equation}
where $\mathfrak{g}$ is the dynamical Lie algebra induced by the PQC, $\mathfrak{g}_j\subseteq \mathfrak{g}$ are simple ideals of $\mathfrak{g}$ according to the decomposition of $\mathfrak{g}$ into a direct sum of commuting ideals:
\begin{equation}\label{eq:DLA decomposition}
    \mathfrak{g} = \mathfrak{g}_1 \oplus \mathfrak{g}_2\oplus\cdots\oplus\mathfrak{g}_k
\end{equation}
where $\mathfrak{g}_i\subseteq \mathfrak{g}$, $i<k$ are simple ideals, and $\mathfrak{g}_k\subseteq \mathfrak{g}$ is the center $Z(\mathfrak{g})$.

The terms $\mathcal{P_{\mathfrak{g}_j}}(\rho)$,  $\mathcal{P_{\mathfrak{g}_j}}(O)$ in Equation \ref{eq:ragone-app} are the $\mathfrak{g}_j$-purities of $\rho$ and $O$, respectively, which are defined as follows: For a Hermitian matrix $H$ and subalgebra $\mathfrak{g}\in u(2^n)$, the \textit{$\mathfrak{g}$-purity} $\mathcal{P}_\mathfrak{g}(H)$ of $H$ is defined as follows:
\[
\mathcal{P}_\mathfrak{g}(H)=
\sum_{j=1}^{\dim\mathfrak{g}}\left|\text{Tr}(B_j^\dagger H)\right|^2
\]
where $\{B_j\}$ is an orthonormal basis of the complexification $\mathfrak{g}_\mathbb{C}$ of $\mathfrak{g}$.

Now, for the main result in Equation \ref{eq:ragone-app}, the circuit ensemble $(\mathcal{E}_L, \nu)$ must form a 2-design over the dynamical Lie group $G = e^{\mathfrak{g}}$, where $\mathfrak{g}$ is the dynamical Lie algebra induced by the PQC. The
required circuit depth $L$ for an $\varepsilon$-approximate 2-design scales as
\begin{equation}
  L \geq \frac{\log(1/\varepsilon)}{\log(1/\|\mathcal A^{(2)}_{\mathcal E_1}\|_\infty)},
  \label{eq:ragone-depth}
\end{equation}
where $\|\mathcal A^{(2)}_{\mathcal E_1}\|_\infty$ is the maximum singular value of the one-layer expressivity superoperator:
\begin{eqnarray}
    \mathcal{A}_{\mathcal{E}_L}^{(2)}H &&= 
    \int_{\mathcal{E}_L}(U\otimes U)H(U^\dagger\otimes U^\dagger) d\mu(U)\nonumber\\
    && -\int_{G}(U\otimes U)H(U^\dagger\otimes U^\dagger) d\mu(U)
\end{eqnarray}
\smallskip
for any $H\in\mathfrak{gl}(\mathcal{H}\otimes\mathcal{H})$, where $\mathcal{H}=(\mathbb{C}^{2})^{\otimes N}$ is the Hilbert space. The deviation from the 2-design approximation is bounded by
\begin{equation}
  \bigl|\mathrm{Var}_{\mathcal{E}_L}[\ell] - \mathrm{Var}_G[\ell]\bigr|
    \leq 3\|\mathcal{A}^{(2)}_{\mathcal{E}_1}\|_\infty^L \,\|O\|_1^2,
  \label{eq:ragone-bound}
\end{equation}
where $\|O\|_1$ is the Schatten 1-norm. Empirically, it is not uncommon to encounter that $\|O\|_1$ grows \emph{exponentially} with $N$ (this occurred for our observations of QAOA on MIS and can be easily verified with simple calculations writing the cost Hamiltonian for QAOA on MIS in terms of Pauli strings), meaning Eq.\ \eqref{eq:ragone-bound}
provides a vacuous bound unless $L$ also grows sufficiently quickly to allow the decay from the infinity-norm term to keep up with the 1-norm growth of our observable. Moreover, computing
$\|\mathcal A^{(2)}_{\mathcal E_1}\|_\infty$ requires enumerating the DLA, which is
$\mathcal{O}(\mathrm{poly}(2^N))$ in the worst case \cite{allcock_dynamical_2024, larocca_diagnosing_2022, wiersema_classification_2023}.

\section{DLA Dimension Empirics}
\label{app:dla-dimension-empirics}
\begin{table}[ht]
\centering
\caption{Summary statistics for DLA dimensions by graph order \(N\). For each \(N\), we report the median DLA dimension and the fraction \(\mathrm{frac}_{\max}\) of graphs attaining the maximal DLA dimension \(4^N\), both for the full graph ensemble and for the vertex-transitive subclass.}
\label{tab:dla-summary}
\bigskip
\begin{tabular}{cccccc}
\toprule
 & \multicolumn{2}{c}{All graphs} & & \multicolumn{2}{c}{Vertex-transitive} \\
\cmidrule{2-3} \cmidrule{5-6}
\(N\) & Median & \(\mathrm{frac}_{\max}\) & & Median & \(\mathrm{frac}_{\max}\) \\
\midrule
1 & 3     & 0     & & 3     & 0    \\
2 & 3.5   & 0     & & 3.5   & 0    \\
3 & 13    & 0     & & 11    & 0    \\
4 & 56.5  & 0     & & 18.5  & 0    \\
5 & 181.5 & 0.294 & & 54    & 0    \\
6 & 4096  & 0.697 & & 207.5 & 0.25 \\
7 & 16384 & 1     & & 8202    & 0.5   \\
\bottomrule
\end{tabular}
\end{table}

Table \ref{tab:dla-summary} summarizes the empirical distribution of DLA dimensions by graph order for the ensemble of all graphs up to order six (aside roughly ten graphs that went over the maximum allotted runtime of 48 hours per submission on our cluster) with six additional randomly sampled graphs from order seven, and for the vertex-transitive subclass up to a vertex count of six, respectively. It will be important to repeat these empirics with tools adapted to QAOA-MIS that are inspired by those in Mao et al. \cite{mao2025qaoamaxcutbarrenplateausgraphs}, as these computations were highly constrained by the expense of finding the DLA for a given generator set (for example, computing the DLA for just the $N=7$ complement cycle graph took approximately 12 hours on an AMD EPYC Rome 7702 with an algorithm implementation that was heavily intra-graph parallelized across cores). Since the distributions are highly skewed and, in several cases, concentrated near the maximal value, we report medians and the fraction of all graphs of maximal DLA dimension rather than relying on means and standard deviations. 

For the full graph ensemble, the median DLA dimension increases rapidly with \(n\), rising from \(3\) at \(N=1\) to \(4096\) at \(N=6\), with the current \(n=7\) sample yielding median \(16384\). More structurally, the fraction of graphs attaining the maximal possible DLA dimension \(4^N\) is zero for \(N \le 4\), becomes nonzero at \(N=5\), and then increases sharply to approximately \(0.66\) at \(N=6\). The current \(N=7\) sample is fully saturated, although this last observation should be regarded as preliminary because the sample size is still small relative to the overall count of graphs at that order. Taken together, these data suggest that, for generic graphs, the ensemble increasingly concentrates on maximal-dimensional DLAs as \(N\) grows.

The vertex-transitive subclass exhibits a qualitatively different behavior over the sampled range. Its median DLA dimensions remain substantially smaller than those of the full graph ensemble, and the fraction of maximal-dimensional cases remains low through \(N=6\). Thus, while the full graph ensemble appears to move rapidly toward maximal DLA dimension, the vertex-transitive subclass shows evidence of a slower growth regime, suggesting that symmetry imposes constraints on DLA growth.

These observations are empirical rather than asymptotic. In particular, the present data do not by themselves prove a precise scaling law for either ensemble. Nevertheless, they provide clear evidence for maximal-dimension DLAs for QAOA on MIS becoming typical in the generic case far more rapidly than in the vertex-transitive setting, although vertex-transitive DLAs do still show large growth in $N$. Determining direct average-case scaling here is difficult since vertex-transitive graphs are highly limited in graph count until much larger $N$, so an asymptotic power-law behavior may still be possible once graph order is large enough for structural properties to normalize. For intuition about why polynomial scaling could be a reasonable expectation here, see Mao et al. \cite{mao2025qaoamaxcutbarrenplateausgraphs}.

\section{Graph Details}
\label{app:graph-details}

\subsection{Random Graph Models}

The \textbf{Erd\H{o}s-R\'enyi} model $G(n,p)$ places edges independently with probability
$p$ \cite{ErdosRenyi1960} The \textbf{Watts-Strogatz} model begins with a regular ring
lattice and rewires each edge independently with probability $\beta$
\cite{watts_strogatz}, producing graphs with high clustering coefficient and small diameter.
The \textbf{Barab\'asi-Albert} model uses preferential attachment to generate scale-free
graphs with power-law degree distributions \cite{barabasi_albert}.  All graph
generation used NetworkX \cite{networkx}.

\subsection{Select Graph Feature Definitions}

The \textbf{von Neumann entropy} of a graph $G$ with Laplacian $L$ is
$S(G) = -\sum_i \lambda_i \log_2 \lambda_i$,
where $\lambda_i$ are the eigenvalues of the density matrix $\rho(G) \equiv L/\mathrm{tr}(L)$
\cite{braunstein_entropy_2006}.

The \textbf{Haemers bound} \cite{haemers_1979} on the independence number is
$\alpha(G) \leq -n\lambda_1\lambda_n/(d_{\min}^2 - \lambda_1\lambda_n)$,
where $\lambda_1 \geq \cdots \geq \lambda_n$ are adjacency eigenvalues and $d_{\min}$ is the
minimum degree.

The \textbf{betweenness centrality} of vertex $v$ is $b(v) = \sum_{u\neq w\in V\setminus\{v\}} \sigma_{uw}(v)/\sigma_{uw}$,
where $\sigma_{uw}$ is the number of shortest $u$--$w$ paths and $\sigma_{uw}(v)$ counts
those passing through $v$.

The \textbf{harmonic centrality} of $v$ is $h(v) = \sum_{u\neq v} 1/d(u,v)$, with
$1/\infty = 0$ for disconnected pairs.

The \textbf{core number} of $v$ is the largest $k$ such that $v$ belongs to a $k$-core
(maximal subgraph with minimum degree $k$).




\newpage
\begin{minipage}[t]{0.47\textwidth}

    \section{EHM Model Performance}
    \label{Comparison:calls}
    \vspace{0.2cm}
    
    \centering
    \centering
    \includegraphics[width=0.85\columnwidth]{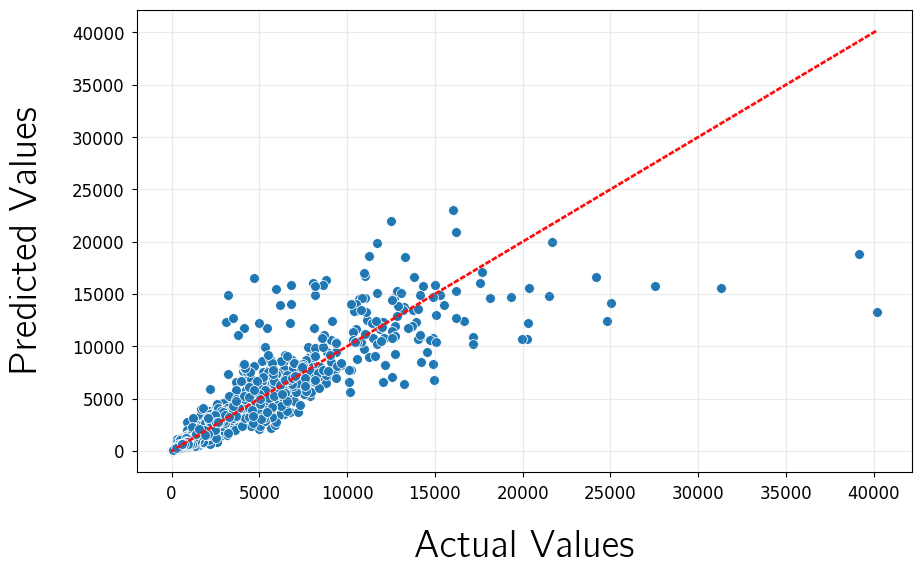}\\
    {\small (a) Random Forest Regression}\\[0.3cm]
    
    \includegraphics[width=0.85\columnwidth]{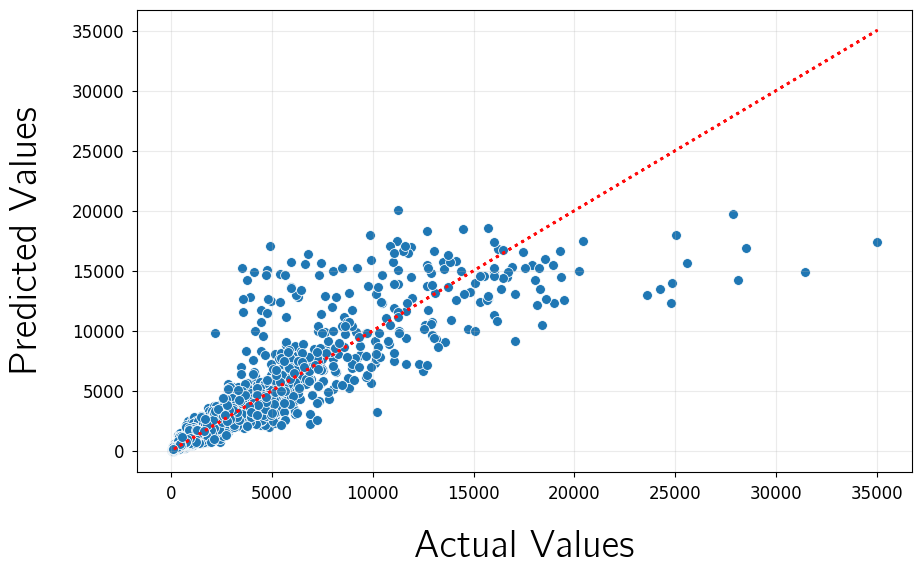}\\
    {\small (b) Neural Network}\\[0.3cm]
    
    \includegraphics[width=0.85\columnwidth]{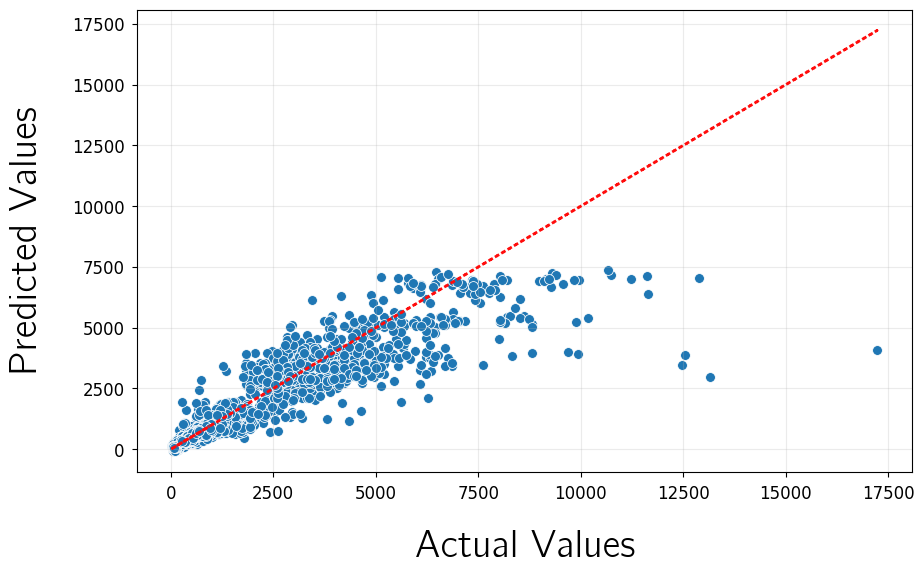}\\
    {\small (c) Graph Neural Network}\\[0.4cm]
    
    {\small\raggedright \noindent \textbf{FIG. 6. $\mathbf{N_{calls}}$ Model Performance.} Performance of empirical hardness models in predicting the number of calls to the quantum circuit: (a) Random Forest Regression, $R^2 = 0.840 \pm 0.016$ (b) Neural Network, $R^2 = 0.823 \pm 0.011$ (c) Graph Neural Network, $R^2 = 0.838\pm 0.017$.\par}
\end{minipage}
\hfill

\begin{minipage}[t]{0.47\textwidth}
    \label{Comparison: mvg}
    \vspace{0.2cm}
    
    \centering
    \centering
    \includegraphics[width=0.85\columnwidth]{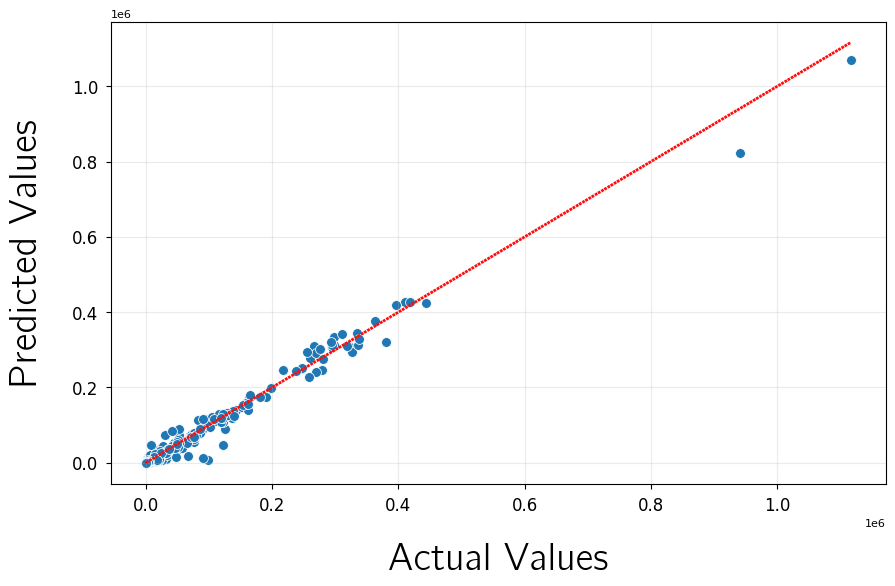}\\
    {\small (a) Random Forest Regression}\\[0.3cm]
    
    \includegraphics[width=0.85\columnwidth]{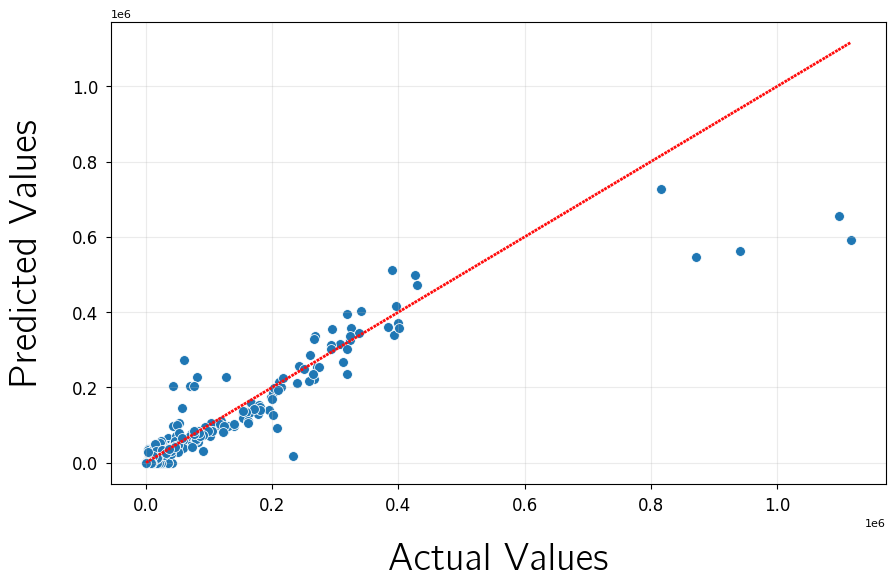}\\
    {\small (b) Neural Network}\\[0.3cm]
    
    \includegraphics[width=0.85\columnwidth]{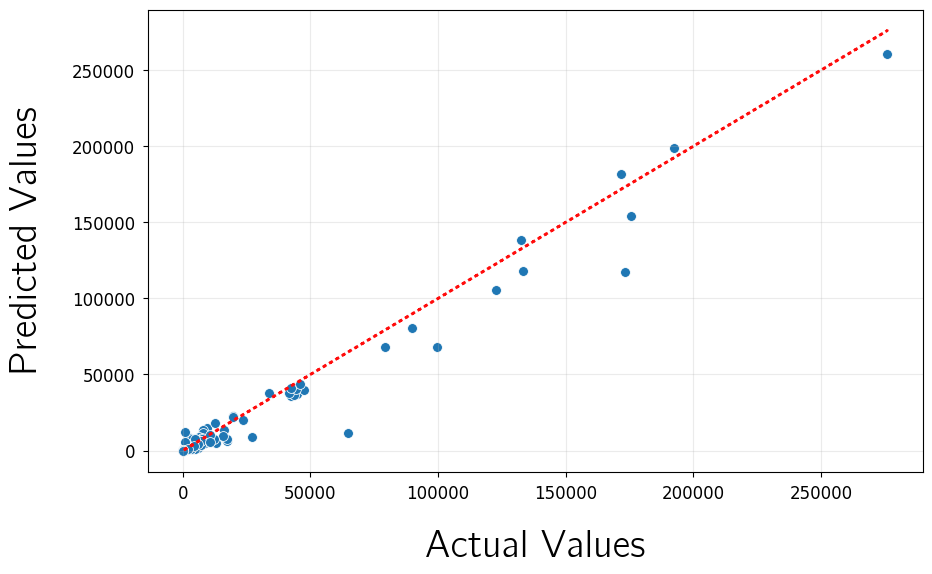}\\
    {\small (c) Graph Neural Network}\\[0.4cm]
    
    {\small\raggedright \noindent \textbf{FIG. 7. Mean Gradient Variance Model Performance.} Performance of empirical hardness models in predicting the mean variance of the gradients. (a) Random Forest Regression, $R^2 = 0.973 \pm 0.010$ (b) Neural Network, $R^2 = 0.837 \pm 0.025$ (c) Graph Neural Network, $R^2 = 0.959\pm 0.009$\par}
\end{minipage}
\hfill

\begin{minipage}[t]{0.47\textwidth}
    \section{Additional EHM Generalization Results}\label{app:generalization}
    \vspace{0.2cm}
    
    \centering
    \centering
    \includegraphics[width=0.85\columnwidth]{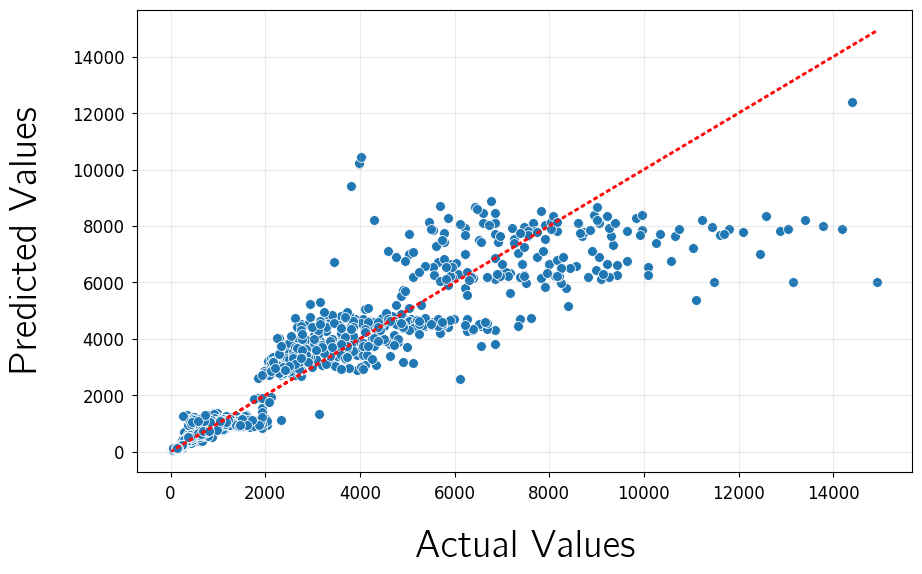}\\
    {\small (a) Random Forest Regression. $R^2 = 0.878\pm 0.002$.}\\[0.3cm]
    
    \includegraphics[width=0.85\columnwidth]{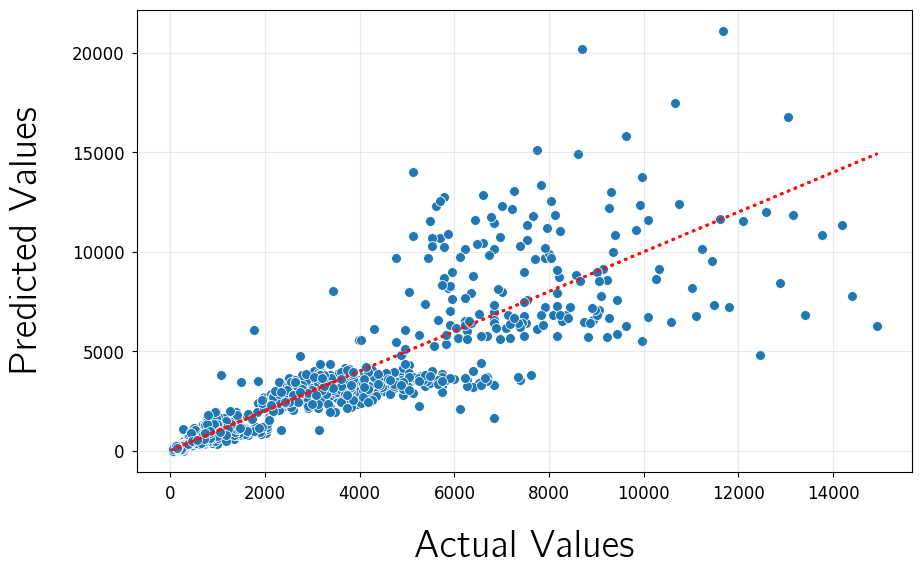}\\
    {\small (b) Neural Network. $R^2 = 0.794\pm 0.036$.}\\[0.3cm]
    
    \includegraphics[width=0.85\columnwidth]{Graphics/gnn_calls_generalize.png}\\
    {\small (c) Graph Neural Network. $R^2 = 0.851\pm 0.006$.}\\[0.4cm]
    
    {\small\raggedright \noindent \textbf{FIG. 8.} Performance of the empirical hardness models when training on N from 1 to 16 and predicting the number of calls to the quantum circuit on N from 19 to 20.\par}
\end{minipage}
\hfill

\begin{minipage}[t]{0.47\textwidth}
    \vspace{0.2cm}
    
    \centering
    \centering
    \includegraphics[width=0.85\columnwidth]{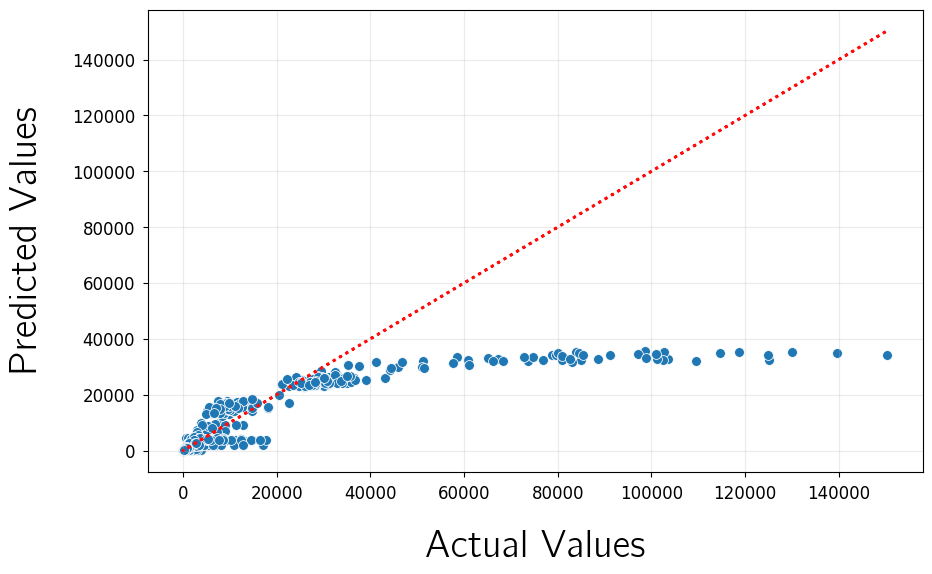}\\
    {\small (a) Random Forest Regression. $R^2 = 0.621\pm 0.004$.}\\[0.3cm]
    
    \includegraphics[width=0.85\columnwidth]{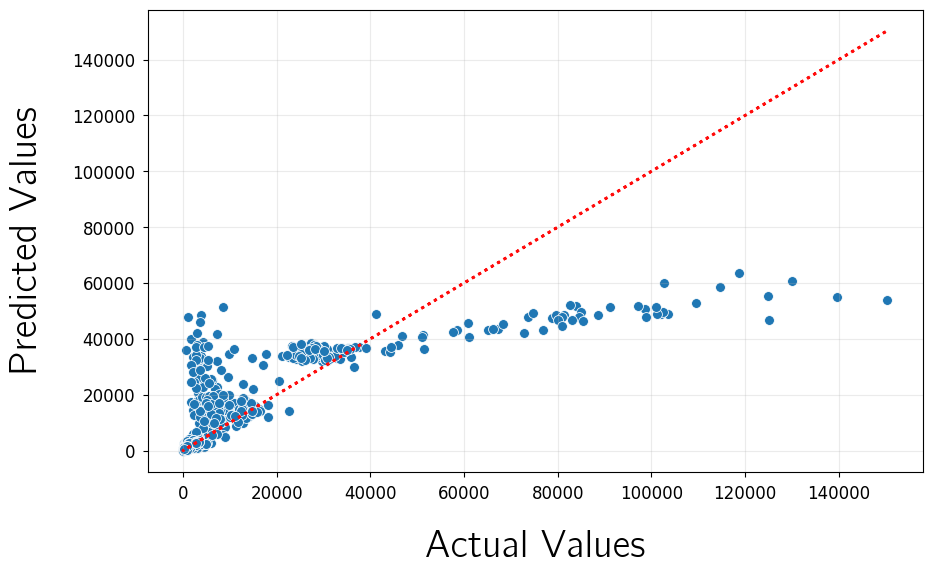}\\
    {\small (b) Neural Network.  $R^2 = 0.647\pm  0.062$. }\\[0.3cm]
    
    \includegraphics[width=0.85\columnwidth]{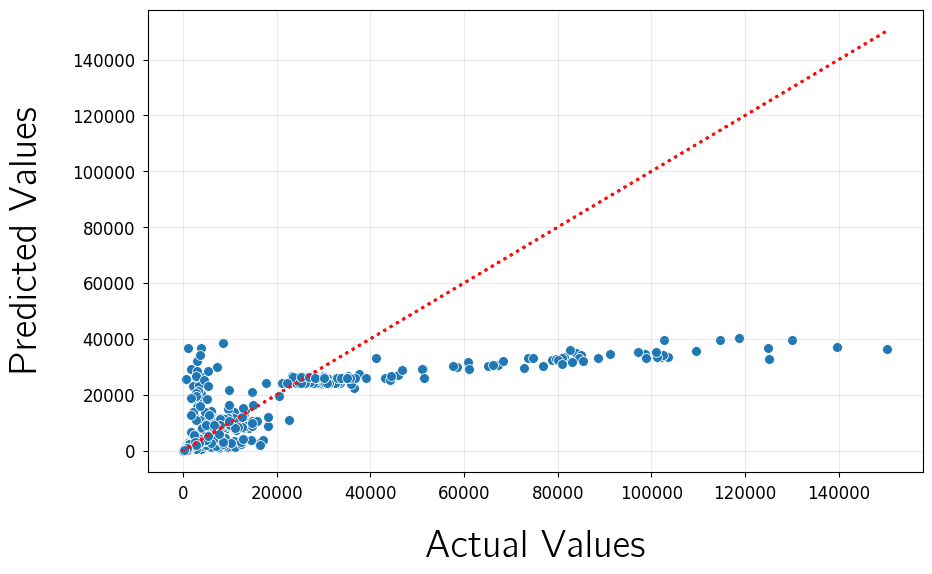}\\
    {\small (c) Graph Neural Network. $R^2 = 0.647\pm  0.062$.}\\[0.4cm]
    
    {\small\raggedright \noindent \textbf{FIG. 9.} Performance of the empirical hardness models when training on N from 1 to 16 and predicting the mean variance of the gradients on N from 19 to 20.\par}
\end{minipage}
\hfill

\end{document}